\documentstyle[A4,12pt,fleqn]{article}
\title{\bf Physical unitarity in the Lagrangian
      {\sl Sp}(2)-symmetric formalism}
\author{{\bf P.M. Lavrov}\thanks{E-mail: lavrov@tspi.tomsk.su.}\\
       {\it Tomsk State Pedagogical University, Tomsk 634041, Russia}\\
       {\bf P.Yu. Moshin}\\
       {\it Tomsk State University, Tomsk 634050, Russia}}
\date{}
\begin{document}
\maketitle
\noindent {\large{\bf Abstract}}\vspace{.5cm}

  The structure of state vector space for a general (non-anomalous)
  gauge theory is studied within the Lagrangian version of the
  $Sp(2)$-symmetric quantization method. The physical {\it S}-matrix
  unitarity conditions are formulated. The general results are
  illustrated on the basis of simple gauge theory models.
%%%%%%%%%%%%%%%%%%%%%%%%%%%%%%%%%%%%%%%%%%%%%%%%%%%%%%%%%%%%%%%%%%%%%%%%%
\section{Introduction}
 \hspace*{\parindent}
 The majority of advanced field theory models are formulated in terms
 of gauge theories. The manifestly covariant quantization of gauge
 theories is carried out in the Lagrangian formalism with the use of
 functional integration. These methods can be divided into two
 groups depending on whether gauge theories quantization is based
 on the principle of invariance under the BRST
 (Becchi--Rouet--Stora--Tyutin) symmetry [1,~2] or the quantization
 rules are underlied by a realization of the principle of invariance
 under the extended BRST symmetry transformations including, along with
 the BRST transformations, also the so-called antiBRST transformations
 [3,~4].

 For arbitrary gauge theories (general gauge theories), the BRST symmetry
 principle has first been realized within the BV (Batalin--Vilkovisky)
 quantization scheme [5,~6] well-known at present. The same principle
 provides the basis of the method [7] of superfield quantization
 discovered for general gauge theories quite recently.

 The studies of Refs.~[8--10] by Batalin, Lavrov and Tyutin have
 suggested a quantization scheme (the Lagrangian $Sp(2)$-symmetric
 formalism), in which, for general gauge theories, the extended BRST
 symmetry has been manifestly realized. Note that the extended BRST
 symmetry principle also underlies a superfield quantization method
 recently proposed for general gauge theories in Ref.~[11].

 Among the number of questions arising in connection with the
 Lagrangian quantization of gauge theories, two problems are of
 great importance. This is first of all the unitarity problem of
 a theory. Next comes the issue of the dependence of a theory upon the
 gauge. These long-standing problems have been explicitly formulated by
 Feynman [12] (the unitarity problem) and Jackiw [13] (the issue of
 gauge dependence).

 For the Yang--Mills type theories, the problem of gauge dependence has
 been thoroughly studied in Refs.~[12--19] (for more references,
 see Ref.~[20]), and for general gauge theories, in Refs.~[21,~22].
 The studies of Refs.~[21,~22] have proved general theorems on the
 gauge dependence of  both the non-renormalized and renormalized
 Green's functions and the $S$-matrix for general gauge theories in
 arbitrary gauges. The theorems themselves have been proved on general
 assumptions of the absence of anomalies, the use of loop expansions
 and the existence of regularization respecting the Ward identities.
 The latest outburst of interest in the gauge dependence within gauge
 theories has been caused by Ref.~[23], in which the author calculated
 the one-loop effective action for Einstein gravity within a special
 class of background gauges and found the effective action to depend
 upon the gauge on-shell which conflicts with the statements of the
 general theorems on gauge dependence [21,~22]. There have been papers
 either maintaining the result [23] and giving reasons for possible
 gauge dependence in arbitrary non-renormalizable gauge theories [24]
 or expressing a doubt [25] about the applicability of the general
 statements [21,~22], as formal ones, to specific theories (Einstein
 gravity, in particular).  The study of Ref.~[26] carried out the
 calculation of the one-loop effective action for Einstein gravity
 within the class of gauges suggested in Ref.~[23]; it was shown,
 firstly, that the general assumptions applied for the proof of the
 theorems [21,~22] are valid in the particular case, secondly, that the
 one-loop effective action for Einstein gravity does not depend upon
 the gauge on-shell in exact conformity with the theorems [21,~22] and,
 finally, that the gauge dependence asserted in Ref.~[23] results from
 incorrect calculations (this observation is also valid as regards
 Ref.~[24]).

 For the non-renormalized Green's functions and the $S$-matrix, the
 gauge dependence within the Lagrangian $Sp(2)$-symmetric quantization
 scheme was studied in Ref.~[27].

 Turning again to the unitarity problem in quantum gauge theories
 within the Lagrangian formalism, note that for the Yang--Mills type
 theories it was efficiently analyzed in Ref.~[28] by Kugo and Ojima in
 the framework of a formalism discovered by them and based on the study
 of the physical subspace ${\cal V}_{\rm phys}$ of the total state
 vector space ${\cal V}$ with indefinite inner product $<\;|\;>$ (note
 that vector spaces having indefinite inner product are also commonly
 referred to as vector spaces with indefinite metric; see, for
 instance, Ref.~[29]).

 The subspace ${\cal V}_{\rm phys}\equiv\{|\rm phys>\}$ is specified by
 the operator $\hat{Q}_{\rm BRST}$ ($\hat{Q}_{\rm BRST}^{\dagger}=
 \hat{Q}_{\rm BRST}$)
\begin{eqnarray}
 \hat{Q}_{\rm BRST}|\rm phys>=0
\end{eqnarray}
 being the generator of the BRST symmetry transformations and
 possessing an important nilpotency property
\begin{eqnarray}
 \hat{Q}_{\rm BRST}^2=0.
\end{eqnarray}
 In the Yang--Mills type theories, the nilpotency of the
 operator $\hat{Q}_{\rm BRST}$ follows immediately from the nilpotency
 of the BRST transformations.

 Even though in arbitrary gauge theories the algebra of the BRST
 transformations is generally open (off-shell), one can still prove (on
 the assumption of the absence of anomalies) that within such theories,
 for the corresponding operator $\hat{Q}_{\rm BRST}$ the nilpotency
 property holds [30]. Thus, one can assume that the Noether charge
 operator $\hat{Q}_{\rm BRST}$ in the BV quantization scheme satisfies
 Eq.~(1.2) and that the Kugo--Ojima formalism, discovered for the
 Yang--Mills type theories, applies to the analysis of the unitarity
 problem for general gauge theories.

 In discussing the property (1.2), it is important to bear in mind that
 the widespread opinion that the nilpotency of the operator
 $\hat{Q}_{\rm BRST}$ guarantees the unitarity of a theory (see, for
 example, Ref.~[31]) proves to be incorrect [32], and that a more
 accurate examination of physicality conditions fulfilment ensuring the
 unitarity of a theory is then required. To this end, we shall now recall
 the main results of analysis of the unitarity problem within the
 framework of the formalism proposed by Kugo and Ojima.

 In Ref.~[28] it was shown that if a theory satisfies the following
 conditions (physicality criteria) for the Hamiltonian $\hat{H}$ and
 the physical subspace ${\cal V}_{\rm phys}$ in the total state vector
 space ${\cal V}$ with indefinite inner product $<\;|\;>$

 \vspace{.2cm}
 (i)
\begin{minipage}[t]{14truecm}
     hermiticity of the Hamiltonian $\hat{H}=\hat{H}^\dagger$ (or
     (pseudo-)unitarity of the total
     $S$-matrix $S^\dagger S=SS^\dagger=1$),
\end{minipage}

 (ii)
\begin{minipage}[t]{14truecm}
 invariance of ${\cal V}_{\rm phys}$ under the time development
 (or $S{\cal V}_{\rm phys}=S^{-1}{\cal V}_{\rm phys}=$
 ${\cal V}_{\rm phys}$)
\end{minipage}
 \hfill(1.3)

 (iii)
\begin{minipage}[t]{14truecm}
 positive semi-defineteness of inner product $<\;|\;>$ in
 ${\cal V}_{\rm phys}$ (${\cal V}_{\rm phys}\ni|\psi>:$ $<\psi|\psi>
 \geq0$),
\end{minipage}
 \vspace{.2cm}
 \\then the physical $S$-matrix $S_{\rm phys}$ is consistently defined
 in a Hilbert space $H_{\rm phys}$ equipped with positive definite inner
 product (the probabilistic interpretation of the quantum theory thus
 secured). Namely, $H_{\rm phys}$ can be identified with a (completed)
 quotient space
\setcounter{equation}{3}
\[
 {\cal V}_{\rm phys}/{\cal V}_0\ni|\widetilde{\Phi}>,\;\;
 |\widetilde{\Phi}>= |\Phi>+{\cal V}_0,\;\;|\Phi>\in{\cal V}_{\rm phys}
\]
 of ${\cal V}_{\rm phys}$ with respect to the zero-norm subspace
 ${\cal V}_0$
\[
 {\cal V}_0=\{|\chi>\in{\cal V}_{\rm
 phys}:<\chi|\chi>=0\},\;\; {\cal V}_{\rm phys}\perp{\cal V}_0,
\]
 where positive definite inner product in ${\cal V}_{\rm phys}/{\cal
 V}_0$ is defined by $<\widetilde{\Phi}|\widetilde{\Psi}>=<\Phi|\Psi>$.
 Given this, for the physical $S$-matrix in $H_{\rm phys}$
\[
 H_{\rm phys}=\overline{{\cal V}_{\rm phys}/{\cal V}_0},\;\;S_{\rm
 phys}|\widetilde{\Phi}>=\widetilde{S|\Phi}>
\]
 the unitarity property holds
\[
 S_{\rm phys}^\dagger S_{\rm phys}=S_{\rm phys}S_{\rm phys}^\dagger=1.
\]
 In this connection, note first of all that the subsidiary condition
 (1.1) ensures, on the assumption of hermiticity of the Hamiltonian, the
 fulfilment of the condition (1.3), (ii) of invariance
 ${\cal V}_{\rm phys}$ under the time development
 (${\cal V}_{\rm phys}^{\rm in}={\cal V}_{\rm phys}^{\rm out}$). In
 Ref.~[27], the analysis of the condition (1.3), (iii) for an arbitrary
 theory (1.2) was based on the study of representation of the algebra of
 the operator $\hat{Q}_{\rm BRST}$ and the ghost charge operator
 $i\hat{Q}_C$ ($[\hat{Q}_C,\hat{H}]=0$)
\[
 [i\hat{Q}_C,\hat{Q}_{\rm BRST}]=\hat{Q}_{\rm BRST}
\]
 (the other commutators trivially vanish) in the one-particle subspace
 of the total Fock space ${\cal V}$.

 The one-particle subspace of the theory generally consists of the
 so-called genuine BRST-singlets, singlet pairs, and quartets [28]. By
 definition, the BRST-singlets are introduced as state vectors
 $|k,N>$ ($i\hat{Q}_C|k,N>=N|k,N>$) from the physical subspace
 ${\cal V}_{\rm phys}$ which cannot be represented in the form $|k,N>=
 \hat{Q}_{\rm BRST}|*>$ for any state $|*>$. Here, {\it k} stands for all
 the quantum numbers (exept the ghost one) which specify the state. At
 that, if $N=0$, then these BRST-singlets are called genuine
 ones and identified with physical states having positive norm.
 Meanwhile, if $N\neq 0$, then state vectors ($|k,-N>$, $|k,N>$) from
 the physical subspace (1.1) possess zero norm and form a singlet pair
 with non-vanishing inner product
\[
 <k,-N|k,N>=1.
\]
 Finally, the states ($|k,N>$, $|k,-N>$, $|k,N+1>$, $|k,-(N+1)>$) such
 that
\[
 |k,N+1>=\hat{Q}_{\rm BRST}|k,N>,\;\;|k,-N>=\hat{Q}_{\rm BRST}|k,-(N+1)>,
\]
\[
 <k,-(N+1)|k,N+1>=<k,-N|k,N>=1
\]
 form a quartet.

 The study of Ref.~[28] discovered a general mechanism, called the
 quartet one, by virtue of which any state that belongs to the physical
 subspace ${\cal V}_{\rm phys}$ of the total Fock space and contains
 quartet particles has vanishing norm. At the same time, the condition
 (1.3), (iii) of positive semi-definiteness of inner product $<\;|\;>$
 in ${\cal V}_ {\rm phys}$ is taken over by a requirement [28] of the
 absence of singlet pairs, which thus guarantees the physical
 $S$-matrix unitarity in the Hilbert space $H_{\rm
 phys}=\overline{{\cal V}_{\rm phys}/{\cal V}_0}$.

 In this paper, the physical $S$-matrix unitarity conditions for a
 general (the anomalies out of account) gauge theory are studied within
 the Lagrangian formulation of $Sp(2)$-symmetric quantization. With
 that, the structure of asymptotic space is analysed on the basis of
 representation of the algebra of operators $\hat{Q}^a$, $i\hat{Q}_C$.
 Here, $\hat{Q}^a$ is an $Sp(2)$-doublet of scalar Noether charge
 operators being the generators of the extended BRST symmetry
 transformations. The physical unitarity analysis implies such general
 assumptions [28] as the non-degeneracy of indefinite inner product
 $<\;|\;>$, the absence of spontaneous symmetry breaking ($\hat{Q}^a|0>=
 \hat{Q}_C|0>=0$), the asymptotic completeness of state vector space
 ${\cal V}$.

 The paper is organized as follows. In Section 2 we summarize the key
 points of the Lagrangian $Sp(2)$-symmetric method and discuss the
 algebraic properties of the extended BRST symmetry transformations (as
 well as their generators) for general gauge theories; in this section we
 use the condensed notations suggested by De Witt [33] and the
 designations adopted in Refs.~[8--10]. Section~3 is devoted to the
 study of the general structure of state vector space and to the
 formulation of the physical unitarity conditions. In Section~4 we
 illustrate the application of the proposed physical unitarity analysis
 on the basis of the study of state vector spaces in concrete gauge
 theory models [32,~34] within the Lagrangian $Sp(2)$ quantization
 method.

%%%%%%%%%%%%%%%%%%%%%%%%%%%%%%%%%%%%%%%%%%%%%%%%%%%%%%%%%%%%%%%%%%%%%%%%%
\section{{\sl Sp}(2)-symmetric Lagrangian quantization}
\hspace*{\parindent}
 Let us now bring to mind the key points of the Lagrangian
 $Sp(2)$-symmmetric method [8--10]. To this end, note first of all
 that the quantization of an arbitrary gauge theory within the
 formalism [8--10] involves introducing a complete set of fields
 $\phi^A$ and the set of the corresponding antifields $\phi^*_{Aa}$
 ($a$=1, 2), $\bar{\phi}_A$ (the doublets of antifields $\phi^*_{Aa}$
 play the role of sources of the BRST and antiBRST transformations
 while the antifields $\bar{\phi}_A$ are the sources of the mixed BRST
 and antiBRST transformations) with the following distribution of the
 Grassmann parity
\[
 \varepsilon(\phi^A)\equiv\varepsilon_A,\;\;
 \varepsilon(\phi^*_{Aa})=\varepsilon_A+1,\;\;
 \varepsilon(\bar{\phi}_A)=\varepsilon_A
\]
 and the ghost number
\[
 {\rm gh}(\phi^*_{Aa})=(-1)^a-{\rm gh}(\phi^A),\;\;
 {\rm gh}(\bar{\phi}_A)=-{\rm gh}(\phi^A).
\]
\hspace*{\parindent}
 The specific structure of configuration space of the fields
 $\phi^A$ (including the initial classical fields, the ghosts, the
 antighosts and the Lagrangian multipliers) is determined by the
 properties of original classical theory, i.e. by the linear
 dependence (reducible theories) or independence (irreducible theories)
 of generators of gauge transformations.  Namely, the studies of
 Refs.~[8,~9] have shown that the fields $\phi^A$ form components of
 irreducible completely symmetric $Sp(2)$-tensors.
 The basic object of the Lagrangian $Sp(2)
 $-symmmetric scheme is the
 bosonic functional $S=S(\phi,\phi^*_{a},\bar{\phi})$, which
 enables one to construct the generating functional of Green's
 functions and satisfies the equations [8--10]
\setcounter{equation}{0}
\begin{eqnarray}
 \frac{1}{2}(S,S)^a+V^aS=i\hbar\Delta^aS
\end{eqnarray}
with the boundary condition
\begin{eqnarray}
 S|_{\phi^{*}_{a}=\bar{\phi}=\hbar=0}={\cal S},
\end{eqnarray}
 where $\cal S$ is the initial gauge invariant classical action. In
 Eq.~(2.1), $\hbar$ is the Planck constant; $(\; ,\; )^a$ is the
 extended antibracket introduced for two arbitrary functionals
 $F=F(\phi,\phi^*_{a},\bar{\phi})$ and $G=G(\phi,\phi^*_{a},\bar{\phi})$
 by the rule
\begin{eqnarray*}
 (F,G)^a=\frac{\delta F}{\delta\phi^A}\frac{\delta G}{\delta
 \phi^{*}_{Aa}}-\frac{\delta G}{\delta\phi^A}\frac{\delta F}
 {\delta\phi^{*}_{Aa}}(-1)^{(\varepsilon(F)+1)(\varepsilon(G)+1)}
\end{eqnarray*}
 (derivatives with respect to the fields are understood as the
 right-hand, and those to the antifields as the left-hand ones), while
 $V^a$ and $\Delta^a$ are operators of the form
\begin{eqnarray*}
 \Delta^a=(-1)^{\varepsilon_A}\frac{\delta_l}{\delta\phi^A}\frac
 {\delta}{\delta\phi^{*}_{Aa}}\;,\;\;
 V^a=\varepsilon^{ab}\phi^{*}_{Ab}\frac{\delta}{\delta\bar\phi_A}\;,
\end{eqnarray*}
 where $\delta_l/\delta\phi^A$ denotes the left-hand derivatives with
 respect to the filds $\phi^A$, and $\varepsilon^{ab}$ is a constant
 antisymmetric second rank tensor of the group $Sp(2)$ subject to the
 normalization condition $\varepsilon^{12}=1$. Note that by virtue of the
 explicit form of the operators $\Delta^a$, $V^a$, the generating equations
 (2.1) can be represented in the form of linear differential equations
\[
 \bar{\Delta}^a\exp\bigg\{\frac{i}{\hbar}S\bigg\}=0,\;\;
 \bar{\Delta}^a=\Delta^a+\frac{i}{\hbar}V^a,
\]
 which enables one, in particular, to establish the compatibility [8]
 of Eq.~(2.1).

 The algebra of operators $\Delta^a$, $V^a$ and the properties of the
 extended antibracket were studied in detail in Ref.~[8], and we omit
 here the corresponding discussion.

 The study of Ref.~[10] proved the existence theorem for solutions of
 Eq.~(2.1) with the boundary condition (2.2) in the form of expansions
 in $\hbar$ powers and described the characteristic arbitrariness of
 solutions.

 The generating functional $Z(J)$ of Green's functions for the fields of
 complete configuration space is constructed, within the Lagrangian
 $Sp(2)$-symmetric quantization scheme, by the rule
\begin{equation}
 Z(J)=Z(J,\phi^*_{a},\bar{\phi})|_{\phi^*_{a}=\bar{\phi}=0},
\end{equation}
 where the extended functional $Z(J,\phi^*_{a},\bar{\phi})$ of
 Green's functions is defined in the form of a functional integral [8]
\begin{equation}
 Z(J,\phi^*_{a},\bar{\phi})=\int\;d\phi\exp\bigg\{\frac{i}{\hbar}\bigg(
 S_{\rm ext}(\phi,\phi^*_{a},\bar{\phi})+J_A\phi^A\bigg)\bigg\}.
\end{equation}
 In Eq.~(2.3), (2.4) $J_A$ are the conventional sourses to the fields
 $\phi^A$ ($\varepsilon(J_A)=\varepsilon_A$, ${\rm gh}(J_A)=-{\rm
 gh}(\phi)$), while $S_{\rm ext}=S_{\rm ext}(\phi,\phi^*_{a},\bar{\phi})$
 is a bosonic functional given by
\begin{equation}
 \exp\bigg\{\frac{i}{\hbar}S_{\rm ext}\bigg\}=\exp\bigg(-i\hbar
 \hat{T}(F)\bigg)\exp\bigg\{\frac{i}{\hbar}S\bigg\},
\end{equation}
 where $S=S(\phi,\phi^*_{a},\bar{\phi})$ is a solution of Eq.~(2.1)
 with the boundary condition (2.2), and $\hat{T}(F)$ is an operator
 defined as
\begin{equation}
 \hat{T}(F)=\frac{1}{2}\varepsilon_{ab}[\bar{\Delta}^b,[\bar{\Delta}^a,
 F]_{-}]_{+}.
\end{equation}
 Here, $F=F(\phi)$ is a bosonic functional fixing a concrete choice of
 admissible gauge, i.e. chosen so as the functional $S_{\rm ext}=
 S_{\rm ext}(\phi,\phi^*_{a},\bar{\phi})$ be non-degenerate in $\phi$
 (examples of such functionals $F$ have been given in Refs.~[8,~9]). It
 follows from the algebraic properties of the operators $\bar{\Delta}^a$
 ($[\bar{\Delta}^a,\bar{\Delta}^b]_{+}=0$) that the functional
 $S_{\rm ext}$ (2.5) satisfies Eq.~(2.1). Note that the gauge fixing
 (2.5) is in fact a particular case of the transformation generated by
 $\hat{T}(F)$, with any bosonic operator chosen for $F$, and describing
 the arbitrariness of solutions of Eq.~(2.1). It should also be pointed
 out that the transformations (2.5) provide a basis for the proof [10]
 of physical equivalence between the quantization of a general gauge
 theory in the BV formalism and the one in the Lagrangian
 $Sp(2)$-symmetric formalism.

 By virtue of the expicit form of the operator $\hat{T}(F)$ (2.6) with
 the functional $F$ depending on the fields of complete configuration
 space only \[ \hat{T}(F)=\frac{i}{\hbar}\frac{\delta F}{\delta\phi^A}
 \frac{\delta}{\delta\bar{\phi}_A}-\frac{1}{2}\varepsilon_{ab}
 \frac{\delta}{\delta\phi^*_{Aa}}\frac{\delta^2F}{\delta\phi^A
 \delta\phi^B}\frac{\delta}{\delta\phi^*_{Bb}}\;,
\]
 the generating functional $Z(J,\phi^*_{a},\bar{\phi})$ of Green's
 functions (2.3), (2.4) is representable in the form [8]
\begin{eqnarray}
 Z(J)&=&\int d\phi\;d\phi^{*}_a\;d\bar{\phi}\;d\lambda\;d\pi^a\;\exp
 \bigg\{\frac{i}{\hbar}\bigg(S(\phi,\phi^{*}_a,\bar{\phi})+\phi^{*}_{Aa}
 \pi^{Aa}+\nonumber\\&&
 +\bigg(\bar{\phi}_A-\frac{\delta F}
 {\delta\phi^A}\bigg)\lambda^A-\frac{1}{2}\varepsilon_{ab}\pi^{Aa}
 \frac{\delta^2F}{\delta\phi^A\delta\phi^B}\pi^{Bb}+
 J_A\phi^A\bigg)\bigg\}\;,
\end{eqnarray}
 where $\pi^{Aa}$, $\lambda^A$
\[
 \varepsilon(\pi^{Aa})=\varepsilon_A+1,\;\;
 {\rm gh}(\pi^{Aa})=-(-1)^a+{\rm gh}(\phi^A),
\]
\[
 \varepsilon(\lambda^A)=\varepsilon_A,\;\;{\rm gh}(\lambda^A)=
 {\rm gh}(\phi^A)
\]
 are auxiliary variables introducing the gauge.

 The validity of Eq.~(2.1) for the functional $S=
 S(\phi,\phi^{*}_a,\bar{\phi})$ enables one, firstly, to establish an
 important fact that the integrand in Eqs.~(2.7) is invariant for $J=0$
 under the following global supersymmetry transformations
\begin{eqnarray}
 \delta\phi^A=\pi^{Aa}\mu_a\,,\;\;\delta\phi^{*}_{Aa}=\mu_a\frac{\delta
 S}{\delta\phi^A}\,,\;\;\delta\bar\phi_A=
 \varepsilon^{ab}\mu_a\phi^{*}_{Ab}\,,\nonumber\\
\end{eqnarray}
\[
 \delta\pi^{Aa}=-\varepsilon^{ab}\lambda^A\mu_b\,,\;\;\delta\lambda^A=0\,,
\]
 where $\mu_a$ is an $Sp(2)$-doublet of constant anticommuting
 infinitesimal parameters. The transformations (2.8) realize the
 extended BRST symmetry transformations in terms of the variables
 $\phi$, $\phi^{*}_a$, $\bar\phi$, $\pi^a$, $\lambda$ and permit
 establishing the independence [8] of the $S$-matrix on a choice of the
 gauge within the Lagrangian $Sp(2)$-symmetric quantization scheme.
 Namely, let us denote the vacuum functional as $Z(0)\equiv Z_F$
 and change the gauge $F\to F+\Delta F$. Then, making in the functional
 integral for $Z_{F+\Delta F}$ the change of variables (2.8) and choosing
 for the parameters $\mu_a$
\begin{eqnarray*}
 \mu_a = \frac {i}{2\hbar}\varepsilon_{ab}
 \frac {\delta(\Delta F)}{\delta \phi^A}\pi^{Ab}\;,
\end{eqnarray*}
 we find that $Z_{F+\Delta F}=Z_F$ and conclude that the $S$-matrix is in
 fact gauge-invariant.

 Secondly, by virtue of Eqs.~(2.1), (2.5), the extended generating
 functional $Z(J,\phi^*_{a},\bar{\phi})$ of Green's functions satisfies
 the Ward identities of the form [8]
\begin{eqnarray}
 \bigg(J_A\frac{\delta}{\delta\phi^*_{Aa}}-\varepsilon^{ab}\phi^*_{Ab}
 \frac{\delta}{\bar\phi_A}\bigg)Z(J,\phi^*_{a},\bar{\phi})=0.
\end{eqnarray}
\hspace*{\parindent}
 The study of Ref.~[27] showed, with the help of Eq.~(2.9), that the
 generating functional $\Gamma=\Gamma(\phi,\phi^{*}_a,\bar\phi)$ of
 vertex functions (derivatives with respect to the sources $J$ are
 understood as the left-hand ones)
\[
 \Gamma(\phi,\phi^{*}_a,\bar\phi)=\frac{\hbar}{i}\ln
 Z(J,\phi^*_{a},\bar{\phi})-J_A\phi^A,\;\;\phi^A=\frac{\hbar}{i}
 \frac{\delta\ln Z(J,\phi^*_{a},\bar{\phi})}{\delta J_A},
\]
 calculated on its extremals $\delta\Gamma/\delta\phi^A=0$, does not
 depend upon the gauge on the hypersurface $\phi^*=0$.

 At the same time, with allowance for Eq.~(2.9), one derives for
 $\Gamma=\Gamma(\phi,\phi^{*}_a,\bar\phi)$ the Ward identities
\begin{eqnarray}
 \frac{1}{2}(\Gamma,\Gamma)^a + V^a\Gamma = 0,
\end{eqnarray}
 which provide a basis for the proof [10] of the $Sp(2)$-invariant
 renormalizability of general gauge theories within the formalism [8--10].

 In particular, Eq.~(2.10), considered at $\phi^{*}_a=\bar\phi=0$,
 results in the invariance of the effective action $\tilde{\Gamma}=
 \tilde{\Gamma}(\phi)$
\begin{eqnarray}
 \tilde{\Gamma}=\Gamma|_{\phi^*_a=\bar{\phi}=0}
\end{eqnarray}
 of the fields $\phi^A$ under the following transformations
\begin{eqnarray}
 \delta\phi^A=\left.\frac{\delta\Gamma}{\delta\phi^*_{Aa}}\right|_
 {\phi^*_a=\bar{\phi}=0}\mu_a
\end{eqnarray}
 (we shall refer to Eq.~(2.12) as quantum extended BRST symmetry
 transformations); namely,
\begin{eqnarray}
 \delta\tilde{\Gamma}=\left.
 \frac{\delta\Gamma}{\delta\phi^A}\frac{\delta\Gamma}
 {\delta\phi^*_{Aa}}\right
 |_{\phi^*_a=\bar{\phi}=0}\mu_a=\left.
 -\varepsilon^{ab}\phi^*_{Ab}\frac{\delta\Gamma}{\delta\bar{\phi}_A}\right
 |_{\phi^*_a=\bar{\phi}=0}\mu_a=0.
\end{eqnarray}
 By virtue of Eq.~(2.10), one readily finds that the algebra of the
 symmetry transformations (2.12), (2.13) is open off-shell
\begin{eqnarray}
 \delta_{(1)}\delta_{(2)}\phi^A&-&\delta_{(2)}\delta_{(1)}\phi^A=
 \nonumber\\
 &=&(-1)^{\varepsilon_A}\left.\frac{\delta\tilde{\Gamma}}{\delta\phi^B}
 \frac{\delta^2\Gamma}{\delta\phi^*_{Bb}\delta\phi^*_{Aa}}\right|_
 {\phi^*_a=\bar{\phi}=0}\mu_{(1)\{a}\mu_{(2)b\}}
\end{eqnarray}
 (here, the symbol $\{\;\}$ denotes the symmetrization with
 respect to the {\it Sp}(2) indices: $A^{\{ab\}}=A^{ab}+A^{ba}$).

 In this connection, note that the study of Ref.~[30] investigated the
 properties of the symmetry transformations $\delta_\alpha$ which form
 an open algebra
\begin{eqnarray}
 \delta_{\alpha}(\delta_{\beta}q^i)-\delta_{\beta}
 (\delta_{\alpha}q^i)=f^{\gamma}_{\alpha\beta}\delta_{\gamma}q^i+
 \Delta_{\alpha\beta}^i
\end{eqnarray}
 within the Lagrangian formulation of an arbitrary non-degenerate
 theory.  Here, $q^i$ are configuration space variables,
 $f^{\gamma}_{\alpha\beta}$ are some structure coefficients (depending
 generally on $q^i$) and $\Delta_{\alpha\beta}^i$ are some functions
 vanishing on-shell. In Ref.~[30] it was shown, on the assumption
 of the absence of anomalies, that within the quantum theory
 constructed in accordance with the Dirac procedure, the following
 relations hold \begin{eqnarray}
 [\hat{Q}_\alpha,\;\hat{H}]=0,\;\;\;[\hat{Q}_{\alpha},\;
 \hat{Q}_{\beta}]=f^{\gamma}_{\alpha\beta}\hat{Q}_{\gamma},
\end{eqnarray}
 where $\hat{H}$ is the Hamiltonian operator and $\hat{Q}_\alpha$ are
 the Noether charge operators generating, on the quantum level, the
 symmetry transformations $\delta_\alpha$.

 The comparison of Eq.~(2.14) with Eqs.~(2.15), (2.16) yields the
 algebra of the operators of Hamiltonian $\hat{H}$ and Noether
 charges $\hat{Q}_{(1)}\equiv\hat{Q}^a\mu_{(1)a}$,
 $\hat{Q}_{(2)}\equiv\hat{Q}^a\mu_{(2)a}$ corresponding to the
 transformations $\delta_{(1)}$, $\delta_{(2)}$ (2.12), (2.14)
\begin{eqnarray}
 [\hat{Q}_{(1,2)},\;\hat{H}]=0,\;\;\;[\hat{Q}_{(1)},\;\hat{Q}_{(2)}]=0.
\end{eqnarray}
 By virtue of the arbitrariness of parameters $\mu_{(1)a}$,
 $\mu_{(2)a}$, Eq.~(2.17) implies the relations
\[
 [\hat{Q}^a,\;\hat{H}]=0,\;\;\;[\hat{Q}^a,\;\hat{Q}^b]_{+}=0.
\]
 Hence it follows that within a general gauge theory (the anomalies out
 of account) there exists a doublet of nilpotent anticommuting
 operators $\hat{Q}^a$ generating the quantum transformations of the
 extended BRST symmetry.

%%%%%%%%%%%%%%%%%%%%%%%%%%%%%%%%%%%%%%%%%%%%%%%%%%%%%%%%%%%%%%%%%%%%%%%%%%%%
\section{Representation of the algebra of {\sl Q}$^a$, {\sl Q}$_C$}
\hspace*{\parindent}
 Let us consider the representation of the algebra
\setcounter{equation}{0}
\begin{eqnarray} [\hat{Q}^a,\hat{Q}^b]_{+}&=&0,\nonumber\\ \\
 {[}i\hat{Q}_C,\hat{Q}^a{]}&=&-(-1)^a\hat{Q}^a\nonumber
\end{eqnarray}
 of the operators $\hat{L}=(\hat{Q}^a,\;i\hat{Q}_C)$ in the
 one-particle subspace ${\cal V}^{(1)}$ of the total Fock space
 ${\cal V}$ with indefinite inner product $<\;|\;>$
\begin{eqnarray}
 \hat{L}{\cal V}^{(1)}\subset{\cal V}^{(1)},\;\;<\Psi|\hat{L}\Phi>=
 <\hat{L}^{\dagger}\Psi|\Phi>,\;\;|\Psi>,\;|\Phi>\in{\cal V}^{(1)},
 \nonumber\\
\end{eqnarray}
\[
 (\hat{Q}^a)^{\dagger}=-(-1)^a\hat{Q}^a,\;\;(\hat{Q}_C)^{\dagger}=
 \hat{Q}_C.
\]
 We shall demonstrate it here that the space ${\cal V}^{(1)}$ of
 representation of the algebra (3.1) is generally a direct sum
\begin{eqnarray}
 {\cal V}^{(1)}=\bigoplus_{n}{\cal V}^{(1)}_{n},\;\;
 \hat{L}{\cal V}^{(1)}_{n}\subset{\cal V}^{(1)}_{n},
 \nonumber\\
\end{eqnarray}
\[
 {\cal V}^{(1)}_{n}\bigcap{\cal V}^{(1)}_{n'}=\emptyset,\;\;n\neq n',
\]
 where subspaces ${\cal V}^{(1)}_{n}$ include the following
 one-particle state complexes

(i) genuine BRST--antiBRST-singlets (physical particles),

(ii) pairs of BRST--antiBRST-singlets,

(iii) BRST-quartets,

(iv) antiBRST-quartets,\hfill (3.4)

(v) BRST--antiBRST-quartets,

(vi) BRST--antiBRST-sextets,

(vii) BRST--antiBRST-octets.

 In order to construct the basis of representation (3.3), (3.4)
 explicitly, note that for an arbitrary state $|\Phi>$ one of the
 following conditions holds
\setcounter{equation}{4}
\begin{eqnarray}
 \frac{1}{2}\varepsilon_{ab}\hat{Q}^a\hat{Q}^b|\Phi>&\neq&0,\\
 \frac{1}{2}\varepsilon_{ab}\hat{Q}^a\hat{Q}^b|\Phi>&=&0.
\end{eqnarray}
 If a state $|\phi_{(k,N)}>\in{\cal V}^{(1)}_{n}$
 ($i\hat{Q}_C|\phi_{(k,N)}>=N|\phi_{(k,N)}>$) satisfies the condition
 (3.5), then, by virtue of Eq.~(3.1), there exists a set of linearly
 independent states
\begin{eqnarray}
 |\phi_{(k,N)}>,\;\hat{Q}^a|\phi_{(k,N)}>,\;
 \frac{1}{2}\varepsilon_{ab}\hat{Q}^a\hat{Q}^b|\phi_{(k,N)}>,
\end{eqnarray}
 which form a basis of a four-dimensional representation of the
 algebra (3.1). Given this, owing to the properties (3.1), (3.2), the
 states
\[
 \hat{Q}^a|\phi_{(k,N)}>,\;\;\frac{1}{2}\varepsilon_{ab}
 \hat{Q}^a\hat{Q}^b|\phi_{(k,N)}>
\]
 have vanishing norm, in particular, $|k,N>\equiv\frac{1}{2}
 \varepsilon_{ab}\hat{Q}^a\hat{Q}^b|\phi_{(k,N)}>$
\begin{eqnarray}
 <k,N|k,N>=0.
\end{eqnarray}
 In accordance with Ref.~[28], for an arbitrary one-particle zero-norm
 (3.8) state $|k,N>$ there exists some (one-particle) state $|k,-N>$
 such that
\begin{eqnarray}
 <k,-N|k,N>=1
\end{eqnarray}
 (by virtue of Eq.~(3.2), any states $|k,N>$, $|k',N'>$ can only have
 a non-vanishing inner product $<k',N'|k,N>$ when $N=-N'$). At the
 same time, among all the basis states, the vector $|k,-N>$
 subject to the normalization (3.9) is without the loss of generality
 unique [28]. In fact, in the subspace of linearly independent states
 ($|k,-N>$, $\{|l,-N>\}$) with the properties $<k,-N|k,N>=<l,-N|k,N>=1$,
 one can always choose a basis ($|k,-N>$,
 $\{|\overline{l,-N}>$$\equiv|l,-N>-|k,-N>\}$) such that
 $<\overline{l,-N}|k,N>=0$. Note that, owing to Eqs.~(3.8), (3.9), the
 basis in the subspace of states $|\Psi>=\{|l,N>,$ $l\neq k\}$,
 $<k,N|\Psi>=0$ can always be chosen so as $<k,-N|l,N>=0$. Indeed, in
 order to go over from the basis states $|k,N>$, $\{|l,N>\}$
\[
 <k,N|k,N>=0,\;\;<k,-N|k,N>=1,
\]
\[
 <k,N|l,N>=0,\;\;<k,-N|l,N>=1,\;\;\forall l
\]
 to an equivalent linearly independent set $|\overline{k,N}>$,
 $\{|\overline{l,N}>\}$
\[
 <\overline{k,N}|\overline{k,N}>=0,\;\;<k,-N|\overline{k,N}>=1,
\]
\[
 <\overline{k,N}|\overline{l,N}>=0,\;\;<k,-N|\overline{l,N}>=0,
 \;\;\forall l
\]
 it is sufficient, for example, to identify
\[
 |\overline{k,N}>=|k,N>,\;\;|\overline{l,N}>=|l,N>-|k,N>,\;\;\forall l.
\]
\hspace*{\parindent}
 From Eqs.~(3.8), (3.9) and the hermiticity assignment (3.2) it follows
 that there exists a set of four states
\begin{eqnarray}
 |\bar{\phi}_{(k,-N)}>,\;\hat{Q}^a|\bar{\phi}_{(k,-N)}>,\;
 \frac{1}{2}\varepsilon_{ab}\hat{Q}^a\hat{Q}^b|\bar{\phi}_{(k,-N)}>,
\end{eqnarray}
 which are also linearly independent and form a basis of
 representation of the algebra (3.1). Here, $|\bar{\phi}_{(k,-N)}>$ is
 a state (3.5) chosen from the condition
\begin{eqnarray}
 \frac{1}{2}\varepsilon_{ab}<\bar{\phi}_{(k,-N)}|\hat{Q}^a
 \hat{Q}^b|\phi_{(k,N)}>=1.
\end{eqnarray}
 By virtue of Eq.~(3.11), the state vectors $|\bar{\phi}_a>\equiv
 (|\bar{\phi}_1>$, $|\bar{\phi}_2>)$ satisfying the normalization
 $<\bar{\phi}_1|\hat{Q}^1\phi>=<\bar{\phi}_2|\hat{Q}^2\phi>=1$
 that corresponds to the zero-norm states $\hat{Q}^a|\phi>$ can be
 chosen in the form $|\bar{\phi}_a>=
 \varepsilon_{ba}(\hat{Q}^b)^\dagger|\bar{\phi}>$.
 Note that without the loss of generality, one can assume
\[
  <\bar{\phi}_{(k,-N)}|\phi_{(k,N)}>=0,
\]
 since if there does exists such $\alpha\neq 0$ that
\[
  <\bar{\phi}_{(k,-N)}|\phi_{(k,N)}>=\alpha,
\]
 then one can choose the basis in the subspace (3.7) so as
\[
 |{\phi'}_{(k,N)}>,\;\hat{Q}^a|{\phi'}_{(k,N)}>,\;
 \frac{1}{2}\varepsilon_{ab}\hat{Q}^a\hat{Q}^b|{\phi'}_{(k,N)}>,
\]
\[
 \frac{1}{2}\varepsilon_{ab}<\bar{\phi}_{(k,-N)}|\hat{Q}^a
 \hat{Q}^b|{\phi'}_{(k,N)}>=1,\;\;
  <\bar{\phi}_{(k,-N)}|{\phi'}_{(k,N)}>=0,
\]
 where $|{\phi'}_{(k,N)}>\equiv\alpha^{-1}|\phi_{(k,N)}>$$-\frac{1}{2}
 \varepsilon_{ab}\hat{Q}^a\hat{Q}^b|\phi_{(k,N)}>$.

 Let us consider the conditions of the linear dependence of the whole
 set of states (3.7), (3.10). In fact, let there among the numbers
 ($\beta,\;\beta_a,\;\tilde{\beta},\;\gamma,\;\gamma_a,\;\tilde{\gamma}$)
 be a non-zero one and let ($|\phi_{(k,N)}>\equiv|\phi>$,
 $|\bar{\phi}_{(k,-N)}>\equiv|\bar{\phi}>$)
\begin{eqnarray}
 \beta|\phi>+\beta_a\hat{Q}^a|\phi>+\frac{\tilde{\beta}}{2}
 \varepsilon_{ab}\hat{Q}^a\hat{Q}^b|\phi>+
 %\nonumber\\
 \gamma|\bar{\phi}>+\gamma_a\hat{Q}^a|\bar{\phi}>+\frac{\tilde{\gamma}}{2}
 \varepsilon_{ab}\hat{Q}^a\hat{Q}^b|\bar{\phi}>=0.
\end{eqnarray}
 Eq.~(3.12), in turn, implies, by virtue of Eq.~(3.1), that there exists
 such $\alpha\neq0$ that
\begin{eqnarray}
 \frac{\alpha}{2}\varepsilon_{ab}\hat{Q}^a\hat{Q}^b|\phi_{(k,N)}>=
 \frac{1}{2}\varepsilon_{ab}\hat{Q}^a\hat{Q}^b|\bar{\phi}_{(k,-N)}>.
\end{eqnarray}
 Namely, if $\beta=\gamma=\beta_a=\gamma_a=0$, then from the condition
 $\tilde{\beta}\neq 0$ it follows that $\tilde{\gamma}\neq 0$
 (reversely, $\tilde{\gamma}\neq 0\Rightarrow\tilde{\beta}\neq 0$) with
 $\alpha=\tilde{\beta}\tilde{\gamma}^{-1}$. In the case
 $\exists a:\beta_a\neq 0$ the condition $\beta=\gamma=0$ implies
 $\gamma_a\neq 0$ (similarly, $\gamma_a\neq 0\Rightarrow\beta_a\neq 0$),
 here $\alpha=\beta_a\gamma_a^{-1}$ (no summation).
 Finally, if $\beta\neq 0$ (or, equivalently, $\gamma\neq 0$), then we
 have $\alpha=\beta\gamma^{-1}$. By virtue of Eq.~(3.13), the state
 $|\bar{\phi}_{(k,-N)}>$ can be chosen from the normalization condition
 (3.11) in the form
\begin{eqnarray}
 |\bar{\phi}_{(k,-N)}>=\alpha|\phi_{(k,N)}>.
\end{eqnarray}
 Hence, evidently, $N=0$, and the spaces of representations
 corresponding to the vector sets (3.7), (3.10) coincide. For a set of
 basis vectors we choose, say, (3.7), i.e. ($|\phi_{(k,N=0)}>\equiv|k,0>$)
\begin{eqnarray}
 |k,0>,\;\;\hat{Q}^a|k,0>,\;\;\frac{1}{2}\varepsilon_{ab}\hat{Q}^a\hat{Q}^b
 |k,0>.
\end{eqnarray}
 Given this, owing to Eq.~(3.14), the relation holds
\begin{eqnarray}
 \frac{{\alpha}^*}{2}\varepsilon_{ab}<k,0|\hat{Q}^a\hat{Q}^b|k,0>=1.
\end{eqnarray}
 By virtue of Eq.~(3.16), the set of states (3.15) can be represented
 in the form of both a BRST-quartet ($(\hat{Q}^1)^{\dagger}=\hat{Q}^1$)
 \begin{eqnarray*}
 |k,0>,\;\;|\overline{k,0}>,\;\;|k,1>,\;\;|k,-1>,
\end{eqnarray*}
\begin{eqnarray}
 |k,1>=\hat{Q}^1|k,0>,\;\;|\overline{k,0}>=\hat{Q}^1|k,-1>,
\end{eqnarray}
\begin{eqnarray*}
 <\overline{k,0}|k,0>=<k,-1|k,1>=1
\end{eqnarray*}
 (choosing for $|k,-1>\equiv-\alpha\hat{Q}^2|k,0>$),
 and an antiBRST-quartet ($(i\hat{Q}^2)^{\dagger}=
 i\hat{Q}^2$)
\begin{eqnarray}
 |k,0>,\;\;|\overline{k,0}>,\;\;|k,-1>,\;\;|k,1>,\nonumber
\end{eqnarray}
\begin{eqnarray}
 |k,-1>=i\hat{Q}^2|k,0>,\;\;|\overline{k,0}>=i\hat{Q}^2|k,1>,
\end{eqnarray}
\begin{eqnarray}
 <\overline{k,0}|k,0>=<k,-1|k,1>=1\nonumber
\end{eqnarray}
 (choosing for $|k,1>\equiv-i{\alpha}^*\hat{Q}^1|k,0>$).
 In what follows, we shall refer to the state complexes of the form
 (3.15), (3.16) as BRST--antiBRST-quartets (3.4), (v).

 In case of linear independence, the states (3.7), (3.10) form a
 BRST--antiBRST-octet (3.4), (vii) and can be represented as both a
 pair of BRST-quartets
\begin{eqnarray}
 (|\phi_{(k,N)}>,\;-\hat{Q}^1\hat{Q}^2|\bar{\phi}_{(k,-N)}>,\;
 \hat{Q}^1|\phi_{(k,N)}>,\;-\hat{Q}^2|\bar{\phi}_{(k,-N)}>),
 \nonumber\\ \\
 (|\bar{\phi}_{(k,-N)}>,\;-\hat{Q}^1\hat{Q}^2|\phi_{(k,N)}>,\;
 \hat{Q}^1|\bar{\phi}_{(k,-N)}>,\;-\hat{Q}^2|\phi_{(k,N)}>)
 \nonumber
\end{eqnarray}
 and a pair of antiBRST-quartets
\begin{eqnarray}
 (|\phi_{(k,N)}>,\;\hat{Q}^2\hat{Q}^1|\bar{\phi}_{(k,-N)}>,\;
 i\hat{Q}^2|\phi_{(k,N)}>,\;-i\hat{Q}^1|\bar{\phi}_{(k,-N)}>),
 \nonumber\\ \\
 (|\bar{\phi}_{(k,-N)}>,\;\hat{Q}^2\hat{Q}^1|\phi_{(k,N)}>,\;
 i\hat{Q}^2|\bar{\phi}_{(k,-N)}>,\;-i\hat{Q}^1|\phi_{(k,N)}>).
 \nonumber
\end{eqnarray}
\hspace*{\parindent}
 Further, we shall consider the states (3.7), (3.10),
 (3.11) (i.e. BRST--antiBRST-quartets and octets), provided they do
 exist in a specific theory, as components of the basis in subspace
 ${\cal V}^{(1)}$. The state vectors (3.7), (3.10), (3.11) evidently
 exhaust all the states (3.5). At the same time, by construction,
 linear combinations of the vectors (3.7), (3.10) constitute, by
 construction, a subspace of states $|\Psi>$, having non-degenerate
 inner product ($\forall|\Psi>\neq0$, $\exists|\Psi'>:$
 $<\Psi|\Psi'>\neq0$), which is invariant under the action of the
 operators $\hat{L}$.

 Consider now the states $|\Phi>\in{\cal V}^{(1)}$ which cannot be
 represented as linear combinations of the vectors (3.7), (3.10), (3.11)
 (i.e. those which do not belong to BRST--antiBRST-quartets or octets);
 from the previous treatment it follows immediately that the states
 $|\Phi>$ under consideration satisfy the condition (3.6). Making
 allowance for the properties of the zero-norm vectors from the set
 (3.7), (3.10), a basis in the subspace of states $|\Phi>$ can
 always be chosen so as $<\Psi|\Phi>=0$ ($|\Psi>$ is an arbitrary
 linear combination of BRST--antiBRST-quartet or octet vectors),
 and hence, the states $|\Phi>$
\[
 <\Psi|\hat{L}\Phi>=<\hat{L}^{\dagger}\Psi|\Phi>=0
\]
 form a space of representation of the algebra of the operators
 $\hat{L}$. Given this, the following conditions generally hold
\begin{eqnarray}
 \exists a:\;\;\hat{Q}^a|\Phi>\neq0,\\ \forall a:\;\;\hat{Q}^a|\Phi>=0.
\end{eqnarray}
\hspace*{\parindent}
 Let us first turn ourselves to the states of the form (3.21). For such
 states the condition is valid ($|*>$ implies arbitrary one-particle
 states)
\begin{eqnarray}
 |\Phi>\neq\hat{Q}^a|*>,
\end{eqnarray}
 since otherwise the states $|\Phi>$ under consideration would be some
 linear combinations of the states (3.7). An arbitrary state $|\Phi>$
 (3.6), (3.21), (3.23), in its turn, satisfies one of the three
 conditions

 (i) \ $\hat{Q}^1|\Phi>\neq0$, \ $\hat{Q}^2|\Phi>\neq0$,

 (ii) \ $\hat{Q}^1|\Phi>\neq0$, \ $\hat{Q}^2|\Phi>=0$,\hfill (3.24)

 (iii) \ $\hat{Q}^1|\Phi>=0$, \ $\hat{Q}^2|\Phi>\neq0$.

\setcounter{equation}{24}
 If a state $|\phi_{(k,N)}>$ satisfies the condition (3.24), (i), then,
 by virtue of Eq.~(3.1), there exist linearly independent states
\begin{eqnarray}
 |\phi_{(k,N)}>,\;\;\hat{Q}^a|\phi_{(k,N)}>,
\end{eqnarray}
 which form a basis of a three-dimensional representation of the
 algebra (3.1). At the same time, the states $\hat{Q}^a|\phi>$ (we
 omit, for the sake of brevity, the notations of the quantum numbers)
 have vanishing norm
\[
 <\hat{Q}^1\phi|\hat{Q}^1\phi>=<\hat{Q}^2\phi|\hat{Q}^2\phi>=0.
\]
 From the above relations it follows, with allowance made for
 Eqs.~(3.2), (3.8), (3.9), that there exist three linearly independent
 states
\begin{eqnarray}
 |\phi_a>,\;\;\frac{1}{2}(\hat{Q}^a)^{\dagger}|\phi_a>,
\end{eqnarray}
 where the states $|\phi_a>\neq\hat{Q}^a|*>$, chosen without the loss
 of generality as eigenvectors for the ghost charge operator
 $i\hat{Q}_C$, satisfy the normalization conditions
\begin{eqnarray}
 <\phi_b|\hat{Q}^a\phi>=\delta^a_b
\end{eqnarray}
 (here, $i\hat{Q}_C|\phi_a>= -(N-(-1)^a)|\phi_a>$ and the conditions
 $<\phi_2|\hat{Q}^1\phi>=<\phi_1|\hat{Q}^2\phi>=0$ hold, therefore,
 automatically); at the same time, by virtue of Eqs.~(3.2), (3.27), we
 have
\begin{eqnarray}
 \frac{1}{2}<(\hat{Q}^a)^{\dagger}\phi_a|\phi>=1,\nonumber\\
\end{eqnarray}
\[
 <(\hat{Q}^b)^{\dagger}\phi_b|\hat{Q}^a\phi_a>=0,\;\;<\phi_a|\phi>=0
\]
 (the inequality $<\phi_a|\phi>\neq0$ leads one to the condition
 $\exists a:N=N-(-1)^a$ and, therefore, does not hold for any $N$).
 Owing to Eqs.~(3.27), (3.28), the bases (3.25), (3.26)
 $(|\phi>$, $\hat{Q}^a|\phi>)\equiv|e_i>$, $(|\phi_a>$,
 $\frac{1}{2}(\hat{Q}^a)^{\dagger}|\phi_a>)\equiv|f_i>$
 are dual with respect to each other $<f_i|e_j>=\delta_{ij}$.
 Hence follows the non-degeneracy of bilinear form $<\;|\;>$ defined on
 the pair $X\equiv\{|e_i>\}$, $Y\equiv\{|f_i>\}$ of state spaces
 corresponding to the vector sets (3.25), (3.26). This fact implies
 that in the space $Y$ exists the (unique) representation
 $\hat{L}^\dagger|f_i>= (\hat{L}^\dagger)_{ij}|f_j>$ of the algebra
 (3.1) conjugate to the representation
 $\hat{L}|e_i>=(\hat{L})_{ij}|e_j>$ defined in $X$, i.e.
 $(\hat{L}^\dagger)_{ij}={(\hat{L})}_{ji}^*$. Namely,
\begin{eqnarray}
 (\hat{Q}^a)^{\dagger}|\phi_b>=\frac{1}{2}\delta^a_b
 (\hat{Q}^c)^{\dagger}|\phi_c>,\nonumber\\
\end{eqnarray}
\[
 (\hat{Q}^a)^{\dagger}(\hat{Q}^b)^{\dagger}|\phi_b>=0
\]
 (for the ghost charge operator $i\hat{Q}_C$, the  basis states of the
 subspace $Y$ are by construction eigenvectors).

 Let us show, with Eqs.~(3.1), (3.2) taken into account, that the
 whole set of states (3.25), (3.26)
\begin{eqnarray}
 |\phi>,\;\;\hat{Q}^a|\phi>,\;\;|\phi_a>,\;\;
 \frac{1}{2}(\hat{Q}^a)^{\dagger}|\phi_a>
\end{eqnarray}
 is linearly independent. Indeed, assuming the reverse, i.e.
\[
 \beta|\phi>+\beta_a\hat{Q}^a|\phi>+\gamma^a|\phi_a>+\frac{\gamma}{2}
 (\hat{Q}^a)^\dagger|\phi_a>=0
\]
 (the numbers ($\beta$, $\beta_a$, $\gamma^a$, $\gamma$) not all
 vanishing), one arrives, by virtue of Eq.~(3.6) and the normalization
 conditions (3.27), at the relation
\[
 \exists a:\;<(\hat{Q}^a)^\dagger\phi|\phi>\equiv\alpha^a\neq0
\]
 representable as
\[
 \beta\neq0\Leftrightarrow\exists a:\;\gamma^a\neq0,\;\alpha^a=(-1)^a
 \gamma^a/\beta,
\]
\[
 \beta=\gamma=0,\;\;\gamma\neq0\Leftrightarrow\exists
 a:\;\beta_a\neq0,\;\alpha^a= -\gamma/\beta_a.
\]
 If we now suppose, for example, that $a=1$, then, owing to Eq.~(3.27)
 $(<\hat{Q}^1\phi_1|\phi>=1)$, the eigenvalues of the ghost charge
 operator $i\hat{Q}_C$ that correspond to the states
 $\hat{Q}^1|\phi>$ and $\hat{Q}^1|\phi_1>$
\[
 i\hat{Q}_C|\hat{Q}^1\phi>=(N+1)|\hat{Q}^1\phi>,
\]
\[
 i\hat{Q}_C|\hat{Q}^1\phi_1>=-N|\hat{Q}^1\phi_1>
\]
 must coincide, i.e. $N+1=-N$. In the case $a=2$ we similarly have
 $N-1=-N$ and find that neither condition can be satisfied for
 an integer $N$.

 Note that the states (3.30), (3.27), (3.29) are representable in the
 form of a BRST-quartet
\[
 |\phi>,\;\;\hat{Q}^1|\phi_1>,\;\;\hat{Q}^1|\phi>,\;\;|\phi_1>,
\]
\[
 <\hat{Q}^1\phi_1|\phi>=<\phi_1|\hat{Q}^1\phi>=1
\]
 and a pair of BRST-singlets ($\hat{Q}^2|\phi>,\;|\phi_2>)$
\begin{eqnarray}
 <\phi_2|\hat{Q}^2\phi>=1,\nonumber\\
\end{eqnarray}
\[
 \hat{Q}^1|\hat{Q}^2\phi>=\hat{Q}^1|\phi_2>=0,\;\;
 |\phi_2>\neq\hat{Q}^1|*>,\;\;|\hat{Q}^2\phi>\neq\hat{Q}^1|*>,
\]
 as well as in the form of an antiBRST-quartet
 ($(\hat{Q}^1)^{\dagger}|\phi_1>=(\hat{Q}^2)^{\dagger}|\phi_2>$)
\[
 |\phi>,\;\;-\hat{Q}^2|\phi_2>,\;\;i\hat{Q}^2|\phi>,\;\;i|\phi_2>,
\]
\[
 <i\phi_2|i\hat{Q}^2\phi>=-<\hat{Q}^2\phi_2|\phi>=1
\]
 and a pair of antiBRST-singlets ($\hat{Q}^1|\phi>,\;|\phi_1>)$
\begin{eqnarray}
 <\phi_1|\hat{Q}^1\phi>=1,\nonumber\\
\end{eqnarray}
\[
 \hat{Q}^2|\hat{Q}^1\phi>=\hat{Q}^2|\phi_1>=0,\;\;
 |\phi_1>\neq\hat{Q}^2|*>,\;\;|\hat{Q}^1\phi>\neq\hat{Q}^2|*>.
\]
 We shall refer to the states (3.30), (3.27) as BRST--antiBRST-sextets
 (3.4), (vi) and consider them (supposing they generally exist in
 a theory) as a part of the basis state vectors in subspace ${\cal
 V}^{(1)}$.

 The above considerations imply that the variety of linear combinations
 of BRST--antiBRST-sextet states contain all the states of the form
 (3.24), (i); at the same time, the sextet representations (3.30),
 (3.29) partly include the states (3.24), (ii), (iii), that is to say
\begin{eqnarray}
 (|\phi_1>,\;\;|\phi_2>)\neq\hat{Q}^a|*>,\nonumber\\
\end{eqnarray}
\[
 \hat{Q}^1|\phi_1>=-\hat{Q}^2|\phi_2>\neq0,\;\;
 \hat{Q}^2|\phi_1>=\hat{Q}^1|\phi_2>=0.
\]
 Reversely, any states (3.33) belong to a BRST--antiBRST-sextet
\begin{eqnarray}
 |\phi>,\;\hat{Q}^1|\phi>,\;\hat{Q}^2|\phi>,\;|\phi_1>,\;
 |\phi_2>,\;\hat{Q}^1|\phi_1>=-\hat{Q}^2|\phi_2>,
\end{eqnarray}
 where $|\phi>$ is chosen from the relations
\[
  <\hat{Q}^1\phi_1|\phi>=-<\hat{Q}^2\phi_2|\phi>=1.
\]
 Hence, for the further analysis of representations of the algebra
 (3.1) that contain the states specified by the conditions (3.6),
 (3.21), (3.23), it is sufficient for us to confine ourselves to the
 states of the form (3.24), (ii), (iii) not repreresentable as linear
 combinations of BRST--antiBRST-sextet states (such states belong,
 without the loss of generality, to the subspace orthogonal to
 BRST--antiBRST-sextet states and, therefore, invariant under the
 action of the operators $\hat{L}$). For the states
\begin{eqnarray}
 |\phi>\neq\hat{Q}^a|*>,\;\hat{Q}^1|\phi>\neq 0,\; \hat{Q}^2|\phi>=0,
\end{eqnarray}
\begin{eqnarray}
 |\bar{\phi}>\neq\hat{Q}^a|*>,\;\hat{Q}^2|\bar{\phi}>\neq 0,\;
 \hat{Q}^1|\bar{\phi}>=0
\end{eqnarray}
 under consideration, the following supplementary conditions hold
\begin{eqnarray}
 \hat{Q}^1|\phi>\neq\hat{Q}^2|*>,
\end{eqnarray}
\begin{eqnarray}
 \hat{Q}^2|\bar{\phi}>\neq\hat{Q}^1|*>.
\end{eqnarray}
 Let us show that the violation, for instance, of the condition (3.37)
 leads one to a contradiction. In fact, $|*>$ is not, by definition,
 representable as a linear combination of BRST--antiBRST-sextet states,
 and, consequently, the relation $\hat{Q}^1|\phi>=\hat{Q}^2|*>$ is only
 possible when $|*>$ belongs to the states (3.36), i.e. without the
 loss of generality, one has
\begin{eqnarray*}
 \hat{Q}^1|\phi>=-\hat{Q}^2|\bar{\phi}>.
\end{eqnarray*}
 From the above relation it follows, by virtue of Eqs.~(3.33)--(3.36),
 that the states ($|\phi>$, $|\bar{\phi}>$, $\hat{Q}^1|\phi>=-\hat{Q}^2
 |\bar{\phi}>$) belong to some BRST--antiBRST-sextet (3.34). The
 inequality (3.38) is proved in a similar way. Eqs.~(3.37), (3.38)
 imply, in particular, that the spaces of representations (3.35) and
 (3.36) respectively cannot be transformed into each another by the
 action of the operators $\hat{L}$.

 By repetition of the given above considerations with respect to
 Eqs.~(3.35)--(3.38) we find that the state complexes (3.35), (3.37)
 constitute some BRST-quartets (3.4), (iii)
\begin{eqnarray}
 |\phi>,\;\;|\phi'>,\;\;\hat{Q}^1|\phi>,\;\;\hat{Q}^1|\phi'>,\nonumber
\end{eqnarray}
\begin{eqnarray}
 <\phi'|\hat{Q}^1\phi>=<\hat{Q}^1\phi'|\phi>=1,
\end{eqnarray}
\[
 |\Phi>\equiv (|\phi>,\;|\phi'>),
\]
\[
 \hat{Q}^2|\Phi>=0,\;|\Phi>\neq\hat{Q}^a|*>,\;\hat{Q}^1|\Phi>\neq
 \hat{Q}^2|*>
\]
 ($|\phi>$ (3.39) is orthogonal to all the BRST--antiBRST-sextet states
 and, in particular, to any state $|\psi>:$ $\forall a$,
 $\hat{Q}^a|\psi>\neq0$; hence, $|\phi'>$ also satisfies Eqs.~(3.35),
 (3.37)), representable as well in the form of two pair of
 antiBRST-singlets
\begin{eqnarray}
 (|\phi>,\;\hat{Q}^1|\phi'>),\;\;(|\phi'>,\;\hat{Q}^1|\phi>).
\end{eqnarray}
 Similarly, the states (3.36), (3.38) constitute antiBRST-quartets
 (3.4), (iv)
\begin{eqnarray}
 |\bar\phi>,\;\;|\bar{\phi}'>,\;\;i\hat{Q}^2|\bar\phi>,\;\;
 i\hat{Q}^2|\bar{\phi}'>,\nonumber
\end{eqnarray}
\begin{eqnarray}
 <\bar\phi'|i\hat{Q}^2\bar\phi>=<i\hat{Q}^2\bar{\phi}'|\bar\phi>=1,
\end{eqnarray}
\[
 |\overline\Phi>\equiv (|\bar\phi>,\;|\bar{\phi}'>),
\]
\[
 \hat{Q}^1|\overline\Phi>=0,\;|\overline\Phi>\neq\hat{Q}^a|*>,\;
 \hat{Q}^2|\overline\Phi>\neq\hat{Q}^1|*>
\]
 and BRST-singlet pairs
\begin{eqnarray}
 (|\bar\phi>,\;i\hat{Q}^2|\bar{\phi}'>),\;\;(|\bar{\phi}'>,\;i\hat{Q}^2
 |\bar\phi>).
\end{eqnarray}
 Thus, with allowance for Eqs. (3.25)--(2.32), (3.35)--(3.42), we have
 described the structure of representations containing the states of
 the form (3.6), (3.21), (3.23).

 Finally, we turn to the states $|\Phi>$ not representable as linear
 combinations of the above considered BRST--antiBRST-quartets, sextets,
 octets and states (3.35)--(3.38) (the state vectors just mentioned are
 without the loss of generality all orthogonal to $|\Phi>$). One readily
 finds that these restrictions can only be met by the states
\[
 |\Phi>\equiv\{|\phi_{(k,N)}>\}\neq\hat{Q}^a|*>
\]
 of the form (3.32) $\hat{Q}^a|\Phi>=0$, which, for $N=0$, we shall
 identify, following Ref.~[28], with physical particles
 $|\phi_k>\equiv|\phi_{(k,N=0)}>$ (genuine BRST--antiBRST-singlets (3.4),
 (i))
\begin{eqnarray}
 <\phi_k|\phi_k>=1,\;\;\hat{Q}^a|\phi_k>=0,\;\;
 |\phi_k>\neq\hat{Q}^a|*>.
\end{eqnarray}
 Meanwhile, in the case $N\neq0$ ($<\phi_{(k,N)}|\phi_{(k,N)}>=0$) we
 shall refer to the states $|\Phi>\equiv(|\phi_{(k,-N)}>$,
 $|\phi_{(k,N)}>$)
\begin{eqnarray}
 <\phi_{(k,-N)}|\phi_{(k,N)}>=1,\;\;\hat{Q}^a|\Phi>=0,\;\;
 |\Phi>\neq\hat{Q}^a|*>.
\end{eqnarray}
 as BRST--antiBRST-singlet pairs (2.4), (ii); with that, the
 state complexes (3.43) and (3.44) are orthogonal to each other.

 Thus, taking Eqs.~(3.7)--(3.44) into account, we have in a general
 case described the structure (3.4) of the one-particle state subspace
 ${\cal V}^{(1)}\supset {\cal V}^{(1)}_n$, (${\cal V}^{(1)}_{n}\perp
 {\cal V}^{(1)}_{n'},\;\;n\neq n'$) as a space of representation
 $\hat{L}{\cal V}\subset {\cal V},\;\;\hat{L}=(\hat{Q}^a,\;i\hat{Q}_C)$
 of the algebra (3.1) of the generators $\hat{Q}^a$ of extended BRST
 symmetry transformations and the ghost charge operator $i\hat{Q}_C$.
 By construction, indefinite inner product $<\;|\;>$ is non-degenerate
 in each subspace ${\cal V}^{(1)}_n$ (see the normalization conditions
 (3.11), (3.27), (3.39), (3.41), (3.43), (3.44) for basis vectors),
 while they themselves have no elements in common (${\cal V}^{(1)}_{n}
 \bigcap{\cal V}^{(1)}_{n'}=\emptyset,\;n\neq n'$) and form a direct sum
 (3.3) of representation subspaces.

%%%%%%%%%%%%%%%%%%%%%%%%%%%%%%%%%%%%%%%%%%%%%%%%%%%%%%%%%%%%%%%%%%%%%%%%
\section{Physical unitarity conditions}
\hspace*{\parindent}
 We now consider, with allowance for Eqs.~(3.3), (3.4), (3.7)--(3.44),
 the conditions of the physical {\it S}-matrix unitarity in the Hilbert
 space $H_{\rm phys}=\overline{{\cal V}_{\rm phys}/{\cal V}_0}$, where the
 physical subspace ${\cal V}_{\rm phys}\ni|\rm phys>$ is specified by the
 {\it Sp}(2)-covariant subsidiary condition
\setcounter{equation}{0}
\begin{eqnarray}
 \hat{Q}^a|\rm phys>=0
\end{eqnarray}
 (which evidently ensures the invariance of ${\cal V}_{\rm phys}$
 under the time development). By virtue of Eq.~(4.1), the structure of
 ${\cal V}_{\rm phys}$ has the form
\[
 {\cal V}_{\rm phys}={\cal V}_{\rm phys}^{1}\bigcap{\cal V}_{\rm phys}^{2},
\]
 where
\begin{eqnarray*}
 {\cal V}\supset{\cal V}_{\rm phys}^{1},\;\;\hat{Q}^1{\cal V}
 _{\rm phys}^{1}=0,\\
 {\cal V}\supset{\cal V}_{\rm phys}^{2},\;\;\hat{Q}^2{\cal V}
 _{\rm phys}^{2}=0.\\
\end{eqnarray*}
 In particular, for the zero-norm subspace ${\cal V}_0\subset{\cal V}$
 we have
\begin{eqnarray}
 {\cal V}_0={\cal V}_0^1\bigcap{\cal V}_0^2,\nonumber\\
\end{eqnarray}
\[
 {\cal V}_0^1\subset{\cal V}_{\rm phys}^1,\;\;
 {\cal V}_0^2\subset {\cal V}_{\rm phys}^2.
\]
 The analysis of representations (3.3), (3.4), (3.7)--(3.44) on the basis
 of the quartet mechanism [28] shows that the state vectors from ${\cal
 V}_{\rm phys}$ containing particles of BRST--antiBRST-quartets (3.4),
 (v) and octets (3.4), (vii) (i.e. state complexes simultaneously
 representable as BRST- (3.17), (3.19) and antiBRST- (3.18), (3.20)
 quartets) belong to the zero-norm subspace ${\cal V}_0$ (4.2).
 The remaining unphysical particles (3.4), (ii), (iii), (iv), (vi)
 generally contain BRST- (antiBRST-) singlet pairs (3.31), (3.32),
 (3.40), (3.42), (3.44) and are present in the physical subspace ${\cal
 V}_{\rm phys}$. In this connection, the physical {\it S}-matrix
 conditions within the suggested approach are the reqirements of
 absence of the pointed out unphysical particles, i.e.
 BRST--antiBRST-singlet pairs (3.4), (ii), BRST-quartets (3.4), (iii)
 (antiBRST-singlet pairs (3.40)), antiBRST-quartets (3.4), (iv)
 (BRST-singlet pairs (3.42)) and BRST--antiBRST-sextets (3.4), (vi).

%%%%%%%%%%%%%%%%%%%%%%%%%%%%%%%%%%%%%%%%%%%%%%%%%%%%%%%%%%%%%%%%%%%%%%%%
\section{State vector spaces in antisymmetric tensor field models}
\setcounter{equation}{0}
\hspace*{\parindent}
 In this section, in order to illustrate the general results of the
 paper, we shall study the physical unitarity conditions for two
 simple gauge theory models [32,~34] within the $Sp(2)$-symmetric
 Lagrangian formalism.

 Consider the theory of a non-abelian antisymmetric field $B_{\mu\nu}^p$
 suggested by Freedman and Townsend [34] and described by the action
\begin{equation} {\cal S} = {\cal S}(A_{\mu}^p, B_{\mu\nu}^p) = \int
 {d^4}x \{-\frac{1}{4} {\varepsilon}^{\mu\nu\rho\sigma} G_{\mu\nu}^p
 B_{\rho\sigma}^p + \frac{1}{2} A_{\mu}^p A^{p\mu}\},
\end{equation}
 where $A_{\mu}^p$ is a vector gauge field with the strength
 $G_{\mu\nu}^p={\partial}_\mu A_{\nu}^p-{\partial}_\nu A_{\mu}^p+f^{pqr}
 A_{\mu}^qA_{\nu}^r$ (the coupling constant is absorbed into the
 structure coefficients $f^{pqr}$), and ${\varepsilon}^{\mu\nu\rho\sigma}$
 is a constant comlpetely antisymmetric four-rank tensor
 (${\varepsilon}^{0123}=1$).

 The aciton (5.1) is invariant under the gauge transformations
\begin{equation}
 {\delta}B_{\mu\nu}^p={\cal D}_{\mu}^{pq}{\xi}_{\nu}^q -
 {\cal D}_{\nu}^{pq}{\xi}_{\mu}^q\equiv R_{\mu\nu\alpha}^{pq}{\xi}^
 {q\alpha},\,\,\,\,\,\,\,\,{\delta}A_{\mu}^p = 0,
\end{equation}
 where $ {\xi}_{\mu}^p $ are arbitrary parameters;
 $ {\cal D}_{\mu}^{pq} $ is the covariant derivative with potential
 $ A_{\mu}^p $ ($ {\cal D}_{\mu}^{pq}={\delta}^{pq}{\partial}_{\mu}+
 f^{prq}A_{\mu}^r $). The algebra of the gauge transformations is
 abelian, and the generators $ R_{\mu\nu\alpha}^{pq} $ have at the
 extremals of the action (5.1) the zero-eigenvectors $ Z_{\mu}^{pq}
 \equiv {\cal D}_{\mu}^{pq} $
\begin{equation}
 R_{\mu\nu\alpha}^{pr}Z^{rq\alpha} =
 {\varepsilon}_{\mu\nu\alpha\beta}f^{prq} \frac{\delta {\cal S}}{\delta
 B_{\alpha\beta}^r}\,,
\end{equation}
 which, in their turn, are linearly independent. According to the
 generally accepted terminology, the model (5.1)--(5.3) is an abelian
 gauge theory of first stage reducibility, and its quantization can
 be carried out, for example, within the BV scheme [6] for the theories
 with linearly dependent (reducible) generators. The study of Ref.~[32]
 showed that the application of the rules [6] to the model (5.1)--(5.3)
 leads one to a physically unitary theory, equivalent to the
 principal chiral field model (non-linear $\sigma$ model for $d=4$).
 We should also mention Refs.~[35,~36] devoted to various aspects of
 quantization of the model (5.1)--(5.3) within the standard BRST
 symmetry.

 Next, consider the gauge model [32], in which the set of fields
 $(A_{\mu}^p, B_{\mu\nu}^p)$ is extended on account of a scalar gauge
 field $\omega^p$ with the transformation rule
\begin{equation}
 \delta\omega^p=\partial^\mu\xi^p_\mu\,,
\end{equation}
 and the initial action of the fields $(A_{\mu}^p, B_{\mu\nu}^p,
 \omega^p)$ is chosen as
\begin{equation}
 {\cal S}(A_{\mu}^p,B_{\mu\nu}^p,\omega^p)=
 {\cal S}(A_{\mu}^p, B_{\mu\nu}^p).
\end{equation}
 Here, ${\cal S}(A_{\mu}^p,B_{\mu\nu}^p)$ is defined in Eq.~(5.1).

 The action (5.5) is invariant under the gauge transformations (5.2),
 (5.4). The generators of these transformations are linearly
 independent (irreducible), and their algebra is abelian. At the
 same time, the Lagrangian quantization of the model (5.5), (5.2), (5.4)
 within the standard BRST symmetry fails [32] to provide physical
 unitarity.

 As mentioned above, the study of Ref.~[10] proved the physical
 equivalence between the Lagrangian quantizations of a general gauge
 theory within the standard (BV formalism) and extended ($Sp(2)$-covariant
 formalism) BRST symmetries. In this connection, we shall reveal, with the
 help of the analysis of asimpotic states of the reducible (5.1)--(5.3) and
 irreducible (5.5), (5.2), (5.4) models within the Lagrangian
 $Sp(2)$-symmetric formalism, the reason for the physical unitarity in
 (5.1)--(5.3) and the origin of unitarity violation in (5.5), (5.2), (5.4)
 (see the proof of the physical {\it S}-matrix unitarity of the theory
 (5.1)--(5.3) within the Lagrangian $Sp(2)$-symmetric formalism in
 Ref.~[37]).

 Consider the model (5.5), (5.2), (5.4) in the Lagrangian
 $Sp(2)$-symmetric quantization of irreducible gauge theories [8].
 To this end, note that the manifest structure of complete
 configuration space $\phi^A$ of the theory has the form
\[
 {\phi}^A=(A^{p\mu},B^{p\mu\nu},\omega^p,B^{p\mu},C^{p\mu a}),
\]
 where $C^{p\mu a}$, $B^{p\mu}$ are the $Sp(2)$-doublets
 of the Faddeev--Popov ghosts and the auxiliary fields respectively,
 introduced according to the number of gauge parameters $\xi^p$
 in Eqs.~(5.2), (5.4). The set of antifields $\phi^*_A$, $\bar{\phi}_A$
 corresponding to the fields $\phi^A$ reads explicitly
\[
 \phi_{Aa}^\ast=(A_{p\mu a}^\ast,B_{p\mu\nu a}^\ast,
 \omega^\ast_{pa},B_{p\mu a}^\ast,C_{p\mu a|b}^\ast),
\]
\[
 \overline{\phi}_A=(\overline{A}_{p\mu},\overline{B}_{p\mu\nu},
 \overline{\omega}_p,\overline{B}_{p\mu},\overline{C}_{p\mu a}).
\]
 The Grassmann parities and the ghost numbers of the fields $\phi^A$
 take on the values
\[
 \varepsilon(A^{p\mu})=\varepsilon(B^{p\mu\nu})=\varepsilon(\omega^p)=
 \varepsilon(B^{p\mu})=0,
\]
\[
 \varepsilon(C^{p\mu a})=1,
\]
\[
 {\rm gh}(A^{p\mu})={\rm gh}(B^{p\mu\nu})={\rm gh}(\omega^p)=
 {\rm gh}(B^{p\mu})=0,
\]
\[
 {\rm gh}(C^{p\mu a})=3-2a.
\]
\hspace*{\parindent}
 The solution of the generating equations (2.1) with the boundary
 condition (2.2) for the model under consideration can be found in a
 closed form. In order to avoid the overloading of the following
 relations with an abundance of indices, we shall further omit the
 gauge indices $p$. Then the bosonic functional
 $S=S(\phi,\phi_a^\ast,\bar{\phi})$ for the theory (5.5), (5.2), (5.4)
 can be represented as
\begin{eqnarray*}
 S&=&{\cal S}+\int d^4x\,\{B_{\mu\nu
 a}^\ast({\cal D}^\mu C^{\nu a}- {\cal D}^\nu C^{\mu a})
 +\omega^*_a\partial^\mu C^a_\mu-\nonumber\\&&
 -\varepsilon^{ab}C^*_{\mu a|b}B^\mu+\overline{B}_{\mu\nu}
 ({\cal D}^\mu B^{\nu}-{\cal D}^\nu B^{\mu})+\overline{\omega}
 \partial^\mu B_\mu\},
\end{eqnarray*}
 where ${\cal S}$ is the initial classical action defined in Eqs.~(5.5),
 (5.1); besides, we have used for the fields $A^p\equiv A$,
 $B^p\equiv B$ the notations
\[
 A^pB^p\equiv AB,
\]
\[
 {\cal D}_{\mu}B\equiv\partial_{\mu}B + A_{\mu}\land B,\;\;(A\land B)^p
 = f^{pqr}A^qB^r.
\]
\hspace*{\parindent}
 Consider the generating functional $Z(J)$ of Green's functions
 represented in the form of the functional integral (2.7) and choose
 for the gauge fixing Boson $F=F(\phi)$
\[
 F=\int d^4x\,\{-\frac{1}{4}B_{\mu\nu}B^{\mu\nu}+\frac{1}{2}\omega^2
 -\frac{1}{4}\varepsilon_{ab}C_\mu^aC^{\mu b}\}.
\]
 Integrating in Eq.~(2.7) over the variables $\lambda^A$, $\pi^{Aa}$,
 $\overline{\phi}_A$, $\phi_{Aa}^\ast$, we obtain, for the theory
 concerned, the following representation of the generating functional
 $Z(J)$
\[
 Z(J)=\int d\phi\;\exp\bigg\{\frac{i}{\hbar}\bigg({\cal S}+S_{\rm FP}
 (\phi)+ S_{\rm GF}(\phi)+J_A\phi^A\bigg)\bigg\},
\]
 where
\begin{eqnarray}
 S_{\rm FP}=-\frac{1}{2}\int d^4x\,
 \{{\varepsilon}_{ab}\partial^{\mu}C^a_{\mu}\partial^{\nu}
 +{\varepsilon}_{ab}{\cal D}_{\mu}C^a_{\nu}( {\cal D}^{\mu}
 C^{\nu b} - {\cal D}^{\nu}C^{\mu b})\},\nonumber\\
\end{eqnarray}
\[
 S_{\rm GF}=\int d^4x\,\{({\cal D}^{\nu}B_{\nu\mu}+\partial_{\mu}\omega)
 B^{\mu}-\frac{1}{2}B_{\mu}B^{\mu}\}.
\]
 The application of the Dirac procedure [38] to the quantum action
 ${\cal S}+S_{\rm FP}+S_{\rm GF}$ (5.6) enables one to establish the
 fact that half the constraints of the theory (the constraints are
 all second-class ones) have the form $\pi=0$. According to the theorem
 [39], these momenta and the corresponding conjugate coordinates can be
 eliminated with the help of the constraint equations; the remaining
 (physical) variables form canonical pairs
\[
 (A^i,\pi_{(A)i}=-\frac{1}{2}{\varepsilon}_{oijk}B^{jk}),
\]
\[
 (B^{oi},\pi_{(B)oi} = B_i),
\]
\begin{equation}
 (\omega ,\pi_{(\omega)} = B_o),
\end{equation}
\[
 (C^{ia},\pi_{(C)ia} = {\varepsilon}_{ab}({\cal D}_oC^b_i -
 {\cal D}_iC^b_o)),
\]
\[
 (C^{oa},\pi_{(C)oa} = {\varepsilon}_{ab}\partial^{\mu}
 C^b_{\mu}),
\]
 for which the Dirac superbrackets coincide with the Poisson
 superbackets.

 The quantum action (5.6) of the theory is invariant under the following
 transformations of the extended BRST symmetry
\[
 \delta B^{\alpha\beta}=({\cal D}^{\alpha}C^{\beta a}-{\cal
 D}^{\beta}C^{\alpha a})\mu_a,\;\;\;\delta A^{\alpha} = 0,
\]
\begin{equation}
 \delta\omega=\partial_{\alpha}C^{\alpha a}\mu_a,
\end{equation}
\[
 \delta C^{\alpha a}=-{\varepsilon}^{ab}B^{\alpha}\mu_b,\;\;\;
 \delta B^{\alpha}=0.
\]
 Specifically, the Lagrangian
\begin{eqnarray*}
 {\cal L}&=&-\frac{1}{4}{\varepsilon}^{\mu\nu\rho\sigma}G_{\mu\nu}
 B_{\rho\sigma} + \frac{1}{2} A_{\mu} A^{\mu}
 -\frac{1}{2}{\varepsilon}_{ab}\partial^{\mu}C^a_{\mu}\partial^{\nu}
 C^b_{\nu} - \frac{1}{2}B_{\mu}B^{\mu}-\nonumber\\
 &&-\frac{1}{2}{\varepsilon}_{ab}{\cal
 D}_{\mu}C^a_{\nu}({\cal D}^{\mu}C^{\nu b} - {\cal D}^{\nu}C^{\mu b}) +
 ({\cal D}^{\nu} B_{\nu\mu} + \partial_{\mu}\omega)B^{\mu}
\end{eqnarray*}
 corresponding to the quantum action (5.6) changes under the
 transformations (5.8) by the total derivative $\delta{\cal L}=
 \partial^\nu F_\nu$
\[
 F_\nu=\{-\frac{1}{2}\varepsilon_{\nu\gamma\rho\sigma}
 G^{\rho\sigma}C^{\gamma a}+({\cal D}_{\nu}C_\gamma^a-
 {\cal D}_{\gamma}C_\nu^a)B^\gamma+B_\nu\partial^\gamma
 C_\gamma^a\}\mu_a.
\]
 This implies the conserved Noether currents $J^a_\nu$
\[
 J_\nu\equiv J^a_\nu\mu_a=\frac{\partial{\cal L}}
 {\partial(\partial^\nu\phi)}\delta\phi-F_\nu,\;\;\partial^\nu J_\nu=0
\]
 (the variations $\delta\phi$ of fields are given by Eq.(5.8)), and
 the corresponding Noether charges $Q^a=\int d^3x\,J^a_0$, expressed in
 terms of the physical variables (5.7), have the form
\begin{eqnarray}
 Q^a = \int d^3x\{\frac{1}{2}{\varepsilon}^{oijk}G_{jk}C^a_i +
 {\varepsilon}^{ab}\pi_{(C)ib}\pi_{(B)}^{oi} + {\varepsilon}^{ab}
 \pi_{(C)ob}\pi_{(\omega)}\}.
\end{eqnarray}
 The algebra of the charges $Q^a$ with respect to the Poisson
 superbracket $\{\;,\;\}$ constructed by the canonically conjugate
 variables (5.7) is abelian
\[
 \{Q^a,Q^b\} = 0.
\]
 As is well-known, within the canonical quantization (according to
 Dirac), classical variables correspond to operators subject to
 canonical (anti-)commutation relaitons resulting from the replacement
 of Poisson (super)brackets by (anti-)commutators (with respect to the
 Grassmann parities of the variables) in accordance with the rule
 $[\;,\;]_{^{+}_{-}}=i\{\;,\;\}$. In particular, the Noether charges
 $Q^a$ (5.9) correspond to the operators $\hat {Q}^a$, generating the
 extended BRST symmetry transformations in terms of the operators of
 physical variables
\[
 [\hat {A^i},\hat{Q}^a]=0,\;\;[\hat{\pi}_{(A)i},\hat{Q}^a]=-i
 \varepsilon_{0ijk}(\partial^j\hat{C}^{ka}+\hat{A}^j\wedge
 \hat{C}^{ka}),
\]
\[
 [\hat {B}^{0i},\hat {Q}^a]=i\varepsilon^{ab}\hat{\pi}^i_{(C)b},\;\;
 [\hat {\pi}_{(B)0i},\hat {Q}^a]=0,
\]
\[
 [\hat {\omega},\hat {Q}^a]=i\varepsilon^{ab}\hat{\pi}_{(C)0b},\;\;
 [\hat {\pi}_{(\omega)},\hat {Q}^a]=0,
\]
\[
 [\hat {C}^{ib},\hat{Q}^a]_{+}=i\varepsilon^{ab}\hat{\pi}_{(B)}^{0i},\;\;
 [\hat{\pi}_{(C)ib},\hat{Q}^a]_{+}=i\varepsilon_{0ijk}\delta_b^a
 (\partial^j\hat{A}^{k}+\frac{1}{2}\hat{A}^j\wedge\hat{A}^k),
\]
\[
 [\hat {C}^{0b},\hat{Q}^a]_{+}=i\varepsilon^{ab}\hat{\pi}_{(\omega)},\;\;
 [\hat{\pi}_{(C)0b},\hat{Q}^a]_{+}=0.
\]
 Given this, a direct verification yields
\[
 [\hat{Q}^a,\hat{Q}^b]_{+}=0.
\]
\hspace*{\parindent}
 Consider the asymptotic state space of the model (5.6), (5.8) and
 study its structure. To this end, we confine ourselves to the analysis
 of free {\it in}-fields (assuming the existence of the corresponding
 {\it in}-limits). The quadratic approximation $S^{(0)}$ of the
 quantum action (5.6) reads ($F_{\mu\nu}\equiv\partial_\mu
 A_\nu-\partial_\nu A_\mu$)
\begin{eqnarray}
  S^{(0)}&=&\int
 {d^4}x\{-\frac{1}{4}{\varepsilon}^{\mu\nu\rho\sigma} F_{\mu\nu}
 B_{\rho\sigma}+\frac{1}{2} A_{\mu} A^{\mu}
 -\frac{1}{2}{\varepsilon}_{ab}\partial_{\mu}C^a_{\nu}\partial^{\mu}
 C^{\nu b} - \nonumber\\ &&
 -\frac{1}{2}B_{\mu}B^{\mu}+({\partial}^{\nu}
 B_{\nu\mu}+\partial_{\mu}\omega)B^{\mu}\}.
\end{eqnarray}
 From Eq.~(5.10) follow the equations of motion for the
 {\it in}-fields (the equations for the {\it in}-operators have the
 same form)
\begin{eqnarray}
 \Box B_{\mu\nu}=0,\;\;\Box\omega=0,\;\;\Box C^a_{\mu}=0, \nonumber\\
\end{eqnarray}
\[
 A_\mu=\frac{1}{2}{\varepsilon}_{\mu\nu\rho\sigma}\partial^{\nu}
 B^{\rho\sigma},\;\;
 B_{\mu}=\partial^{\nu}B_{\nu\mu}+\partial_\mu\omega
\]
 (we omit the symbol of {\it in}-limit). The solution of Eq.~(5.11) for
 the operator-valued field $\hat{B}_{\mu\nu}(x)$ is representable as
 ($(\hat{B}^{(-)}_{\mu\nu})^{\dagger}=\hat{B}^ {(+)}_{\mu\nu}$)
\begin{eqnarray}
 \hat{B}_{\mu\nu}(x)=\int\frac{d^3k}{\sqrt{2(2\pi)^3k_0}}
 \bigg(\hat{B}^{(-)}_{\mu\nu}(\vec{k})e^{-ikx}+\hat{B}^{(+)}_{\mu\nu}
 (\vec{k})e^{ikx}\bigg).
\end{eqnarray}
 Similar decompositions  are valid for the operators $\hat{\omega}(x)$,
 $\hat{C}^a_{\mu}(x)$ with allowance made for
\begin{eqnarray}
 (\hat{\omega}^{(-)})^{\dagger}=\hat{\omega}^{(+)},\;\;
 (\hat{C}^{a(-)}_{\mu})^{\dagger}=(-1)^{a+1}\hat{C}^{a(+)}_{\mu}.
\end{eqnarray}
 At the same time, the analysis of equal-time (anti-)commutation
 relations yields
\begin{eqnarray*}
 [\hat{B}^{(-)}_{\mu\nu}(\vec{k}),\hat{B}^{(+)}_{\rho\sigma}(\vec{k'})]
 =\eta_{\mu[\rho}\eta_{\sigma]\nu}\delta(\vec{k}-\vec{k'}),
\end{eqnarray*}
\begin{eqnarray*}
 {[}\hat{\omega}^{(-)}(\vec{k}),\hat{\omega}^{(+)}(\vec{k'}){]}
 =\delta(\vec{k}-\vec{k'}),
\end{eqnarray*}
\begin{eqnarray*}
 {[}\hat{C}^{a(-)}_{\mu}(\vec{k}),\hat{C}^{b(+)}_{\nu}(\vec{k'}){]}_{+}
 ={\varepsilon}^{ab}\eta_{\mu\nu}\delta(\vec{k}-\vec{k'}).
\end{eqnarray*}
 The action (5.10) of the {\it in}-fields is invariant under the
 following (non-vanishing) transformations
\[
 \delta B^{\alpha\beta}=(\partial^\alpha C^{\beta a}-\partial^\beta
 C^{\alpha a})\mu_a,
\]
\[
 \delta\omega=\partial_{\alpha}C^{\alpha a}\mu_a,
\]
\[
 \delta C^{\alpha a}=-{\varepsilon}^{ab}B^{\alpha}\mu_b.
\]
 In terms of the creation operators, the corresponding transformations
 for $\hat{B}_{\mu\nu}$, $\hat{\omega}$, $\hat{C}^a_{\mu}$ have the
 form
\begin{eqnarray*}
 [\hat{B}^{(+)}_{\mu\nu}(\vec{k}),\hat{Q}^{a}]=-\bigg(k_\mu
 \hat{C}^{a(+)}_\nu(\vec{k})-k_\nu\hat{C}^{a(+)}_\mu(\vec{k})\bigg),
\end{eqnarray*}
\begin{eqnarray}
 {[}\hat{\omega}^{(+)}(\vec{k}),\hat{Q}^{a}{]}=-k^\mu\hat{C}^{a(+)}
 _{\mu}(\vec{k}),
\end{eqnarray}
\begin{eqnarray*}
 {[}\hat{C}^{b(+)}_{\mu}(\vec{k}),\hat{Q}^{a}{]}_{+}
 ={\varepsilon}^{ab}\bigg(k_\mu\hat{\omega}^{(+)}(\vec{k})+k^\nu
 \hat{B}^{(+)}_{\nu\mu}(\vec{k})\bigg).
\end{eqnarray*}
 In Eq.~(5.14), $\hat{Q}^{a}$ is a doublet of generators of the
 extended BRST symmetry transformations for the operatorial
 {\it in}-fields, its normal form being
\begin{eqnarray*}
 \hat{Q}^{a}&=&\int d^3k\;k^\mu\bigg(\hat{\omega}^{(+)}(\vec{k})
 \hat{C}^{a(-)}_\mu(\vec{k})+C^{a(+)}_\mu(\vec{k})\hat{\omega}^
 {(-)}(\vec{k})+\\&& \mbox{}+
 \hat{B}^{(+)}_{\mu\nu}(\vec{k})\hat{C}^{\nu a(-)}(\vec{k})+
 \hat{C}^{\nu a(+)}(\vec{k})\hat{B}^{(-)}_{\mu\nu})(\vec{k})\bigg).
\end{eqnarray*}
 The analysis of asymptotic state space structure is conveniently
 carried out with the help of the following local basis
\begin{eqnarray}
 e_L^\mu(\vec{k})&=&\frac{1}{2k^2_0}k^\mu=\frac{1}{2k^2_0}(k^0,\;
 \vec{k}), \nonumber\\
 e_T^\mu(\vec{k})&=&\frac{1}{2k^2_0}\bar{k}^\mu=
 \frac{1}{2k^2_0}(k^0,\;-\vec{k}),\\
 e_\lambda^\mu(\vec{k})&=&(0,\;\vec{\varepsilon}_\lambda(\vec{k})),
 \;\;
 \vec{k}\vec{\varepsilon}_\lambda(\vec{k})=0,\;
 \vec{\varepsilon}_\lambda(\vec{k})\vec{\varepsilon}_{\lambda'}
 (\vec{k})=\delta_{\lambda\lambda'},\;\lambda=1,\;2.\nonumber
\end{eqnarray}
 Given this, the decomposition of any vector $a^\mu(\vec{k})$ in the
 basis vectors (5.15) have the form
\[
 a^\mu(\vec{k})=e_L^\mu(\vec{k})a_L(\vec{k})+
 e_T^\mu(\vec{k})a_T(\vec{k})+
 e_\lambda^\mu(\vec{k})a_\lambda(\vec{k}),
\]
 where
\[
 a_L(\vec{k})=\bar{k}^\mu a_\mu(\vec{k}),\;\;
 a_T(\vec{k})=k^\mu a_\mu(\vec{k}),\;\;
 a_\lambda(\vec{k})=-e_{\lambda}^\mu(\vec{k})a_\mu(\vec{k}).
\]
\hspace*{\parindent}
 Note that for the study of one-particle state space it is
 sufficient to analyze the structure of creation operators. We
 shall decompose all the vector creation operators in the basis (5.15),
 representing the operator $\hat{B}^{(+)}_{\mu\nu}$ as
\begin{eqnarray}
 \hat{B}^{(+)}_{\mu\nu}=\bigg(\hat{B}^{(+)}_{oi},\;\;\frac{1}{2}
 {\varepsilon}_{oijk}\hat{B}^{jk(+)}\equiv\hat{D}^{(+)}_i\bigg).
\end{eqnarray}
 Namely,
\[
 {[}\hat{D}^{(+)}_L(\vec{k})-\hat{D}^{(+)}_T(\vec{k}),\hat{Q}^a]=0,
\]
\[
 {[}\hat{D}^{(+)}_\lambda(\vec{k}),\hat{Q}^a]=-k_0\varepsilon_{\lambda
 \lambda'}\hat{C}^{a(+)}_{\lambda'}(\vec{k}),\;\;
 \varepsilon_{\lambda\lambda'}=-\varepsilon_{\lambda'\lambda},\;\;
 \varepsilon_{12}=-1,
\]
\[
 {[}\hat{B}^{(+)}_L(\vec{k})-\hat{B}^{(+)}_T(\vec{k}),\hat{Q}^a]=
 2k_0\hat{C}^{a(+)}_T(\vec{k}),
\]
\[
 {[}\hat{B}^{(+)}_\lambda(\vec{k}),\hat{Q}^a]=-k_0
 \hat{C}^{a(+)}_{\lambda}(\vec{k}),
\]
\[
 {[}\hat{\omega}^{(+)}(\vec{k}),\hat{Q}^a]=-\hat{C}^{a(+)}_T(\vec{k}),
\]
\[
 {[}\hat{C}^{b(+)}_L(\vec{k}),\hat{Q}^a]_{+}=-2\varepsilon^{ab}
 \{\hat{\omega}^{(+)}(\vec{k})+\frac{1}{2k_0}
 (\hat{B}^{(+)}_L(\vec{k})-\hat{B}^{(+)}_T(\vec{k}))\},
\]
\[
 {[}\hat{C}^{b(+)}_T(\vec{k}),\hat{Q}^a]_{+}=0,
\]
\[
 {[}\hat{C}^{b(+)}_\lambda(\vec{k}),\hat{Q}^a]_{+}=-\varepsilon^{ab}k_0
 \hat{B}^{(+)}_\lambda(\vec{k}).
\]
 Hence it follows that among the 15 ($p$, $\vec{k}$ fixed) one-particle
 states there is only one genuine BRST--antiBRST-singlet:
 $(\hat{D}^{(+)}_L-\hat{D}^{(+)}_T)|0>$. There are two
 BRST--antiBRST-quartets (respectively, for $\lambda=1,\;2$)
\begin{eqnarray*}
 \hat{D}^{(+)}_\lambda|0>,\;\;\hat{Q}^a\hat{D}^{(+)}_\lambda|0>,\;\;
 \frac{1}{2}{\varepsilon}_{ab}\hat{Q}^a\hat{Q}^b\hat{D}^{(+)}_
 \lambda|0>
\end{eqnarray*}
 and a BRST--antiBRST-sextet
\begin{eqnarray*}
 \hat{\omega}^{(+)}|0>,\;\;\hat{Q}^a\hat{\omega}^{(+)}|0>,\;\;
 \hat{C}^{a(+)}_L|0>,\;\;
 \frac{1}{2}{\varepsilon}_{ab}\hat{Q}^a\hat{C}^{b(+)}_L|0>.
\end{eqnarray*}
\hspace*{\parindent}
 The presence of the BRST--antiBRST-sextet in the state space thus
 accounts for the physical {\it S}-matrix unitarity violation [32] in
 the theory concerned.

 Let us now turn ourselves to the quantizaiton of the reducible model
 (5.1)--(5.3) within the Lagrangian $Sp(2)$-symmetric scheme [8--10].

 In accordance with the rules [9], we introduce the set of fields
 $\phi^A$
\[
 {\phi}^A=(A^{\mu}, B^{\mu\nu}, B^{\mu}, B^{a}, C^{\mu a},
 C^{ab})
\]
 and the sets of the corresponding antifields ${\phi}_{Aa}^{\ast}$,
 $\overline{\phi}_A$
\[
 \phi_{Aa}^\ast=(A_{\mu a}^\ast,B_{\mu\nu a}^\ast,B_{\mu a}^\ast,
 B_{a|b}^\ast,C_{\mu a|b}^\ast,C_{a|bc}^\ast),
\]
\[
 \overline{\phi}_A=(\overline{A}_{\mu},\overline{B}_{\mu\nu},
 \overline{B}_{\mu},\overline{B}_{a},\overline{C}_{\mu a},
 \overline{C}_{ab}).
\]
 Note that $C^{ab}$, $B^a$ are the ghost fields
 (symmetric second rank {\it Sp}(2)-tensors) and {\it Sp}(2)-doublets
 of first stage respectively, introduced in accordance with the number of
 gauge parameters $\xi\equiv\xi^p$ for the generators $R_{1\mu\nu}^{pq}
 \equiv R_{\mu\nu\alpha}^{pr}Z^{rq\alpha}$. Given this,
\[
 \varepsilon(C^{ab})=0,\;\;\;{\rm gh}(C^{ab})=6-2(a+b),
\]
\[
 \varepsilon(B^{a})=1,\;\;\;{\rm gh}(B^{a})=3-2a
\]
 (the remaining fields $A^{\mu}$, $B^{\mu\nu}$, $B^{\mu}$, $C^{\mu a}$
 have been described above).

 The solution $S=S(\phi,\phi_a^\ast,\bar{\phi})$ of the generating
 equations (2.1) with the boundary condition (2.2) for the model
 (5.1)--(5.3) can be represented as
\begin{eqnarray*}
 S&=&{\cal S}+\int d^4x\,\{B_{\mu\nu a}^\ast({\cal D}^\mu C^{\nu a}-
 {\cal D}^\nu C^{\mu a})-\varepsilon^{ab}C_{\mu a|b}^\ast B^{\mu}
 +\overline{B}_{\mu\nu}({\cal D}^\mu B^{\nu}-{\cal D}^\nu B^{\mu})+\\&&
 +C_{\mu a|b}^\ast D^\mu C^{ab}-2\varepsilon^{ab}C_{a|bc}^\ast B^c
 -B_{\mu a}^\ast D^\mu B^a+2\overline{C}_{\mu a}D^\mu B^a-\\&&
 -\varepsilon^{\mu\nu\rho\sigma}B_{\mu\nu a}^\ast(\overline
 {B}_{\rho\sigma}\wedge B^a)+\frac{1}{2}\varepsilon^{\mu\nu\rho\sigma}
 (B_{\mu\nu a}^\ast\wedge B_{\rho\sigma b}^\ast)C^{ab}\},
\end{eqnarray*}
 where ${\cal S}$ is the classical action (5.1).

 Choosing for the gauge fixing bosonic functional $F=F(\phi)$
\[
 F=\int d^4x\,\{-\frac{1}{4}B_{\mu\nu}B^{\mu\nu}-\frac{1}{4}
 \varepsilon_{ab}C_\mu^aC^{\mu b}-
 \frac{1}{12}\varepsilon_{ab}\varepsilon_{cd}C^{ac}C^{bd}\}
\]
 and integrating in Eq.~(2.7) over the variables $\lambda$, $\pi^{a}$,
 $\overline{\phi}$, $\phi_{a}^\ast$, one arrives at the generating
 functional $Z(J)$ of Green's functions of the form
\[
 Z(J)=\int d\phi\;\Delta\;\exp\bigg\{\frac{i}{\hbar}\bigg({\cal S}+
 S_{\rm FP}(\phi)+S_{\rm GF}(\phi)+J_A\phi^A\bigg)\bigg\},
\]
 where
\[
 S_{\rm FP}=\int\,d^4x\{\frac{1}{4}G_{\mu\nu}^aM_{ab}K_c^{b[\mu\nu]
   [\rho\sigma]}G_{\rho\sigma}^c-\frac{1}{4}\varepsilon_{ab}
   \varepsilon_{cd}D_\mu C^{ac}D^\mu C^{bd}\},
\]
\begin{equation}
 S_{\rm GF}=\int\,d^4x\{B_\mu D_\nu B^{\nu\mu}+
 \varepsilon_{ab}B^aD_\mu C^{\mu b}-\frac{1}{2}B_\mu B^\mu
 -\frac{1}{2}\varepsilon_{ab}B^aB^b\},
\end{equation}
\[
  \Delta=\int dB_a^\ast\;\exp\bigg\{\frac{2i}{\hbar}\int
  d^4x\,B_{0ib}^\ast M^{bc} B_{0jc}^\ast \eta^{ij}\bigg\}.
\]
 In Eq.~(5.17), we have used the following notations
\[
 K_b^{a[\mu\nu][\rho\sigma]} \equiv \frac{1}{2}\{\delta_b^a
 (\eta^{\mu\rho}\eta^{\nu\sigma}-\eta^{\mu\sigma}\eta^{\nu\rho})+
 X_b^a\varepsilon^{\mu\nu\rho\sigma}\},
\]
\[
 G_{\mu\nu}^a \equiv{\cal D}_\mu C_\nu^a-{\cal D}_\nu
 C_\mu^a-\frac{1}{2}\varepsilon_{\mu\nu\rho\sigma}B^a\wedge
 B^{\rho\sigma}.
\]
 The matrix $M_{ab}$ is the inverse of $M^{ab}$
\[
  M^{ab} \equiv \varepsilon^{ab}-X_c^aX_d^b\varepsilon^{cd},\;\;\;\;
  M^{ac}M_{cb}=\delta_b^a,
\]
 while the action of the matrix $X_b^a$ on the objects $E\equiv E^p$
 carring the gauge indices $p$ is defined by the rule
\[
 X_b^aE \equiv \varepsilon_{bc}(C^{ac}\wedge E).
\]
 The functional $\Delta$ (5.17) can be considered as a contribution to
 the integration measure (in $\phi$ space), invariant under the extended
 BRST symmetry transformations
\begin{eqnarray*}
 \delta B^{\alpha\beta}&=&-\varepsilon^{ab}M_{bc}K_d^{c[\alpha\beta]
 [\gamma\delta]}G_{\gamma\delta}^d\mu_a,\;\;\;
 \delta A^{\alpha}=0,\\
 \delta C^{\alpha a}&=&({\cal D}^\alpha C^{ab}-\varepsilon^{ab}B^
 \alpha)\mu_b,\;\;\;\delta B^\alpha={\cal D}^\alpha B^a\mu_a,\\
 \delta C^{ab}&=&B^{\{a}\varepsilon^{b\}c}\mu_c,\;\;\;
 \delta B^a=0
\end{eqnarray*}
 for the quantum action ${\cal S}+S_{\rm FP}+S_{\rm GF}$ (5.17).
 The corresponding Noether charges $Q^a$
\begin{eqnarray*}
 Q^a&=&\int d^3x\{-\frac{1}{2}{\varepsilon}^{oijk}G_{jk}C^a_i+\pi_{(C)ib}
 {\cal D}^iC^{ab}-{\varepsilon}^{ab}\pi_{(C)0b}{\cal D}^iB_{0i}+\\&&
 +{\varepsilon}^{ab}\pi_{(C)ib}\pi_{(B)}^{0i}+2{\varepsilon}^
 {ab}{\varepsilon}^{cd}\pi_{(C)bc}\pi_{(C)od}\},
\end{eqnarray*}
 expressed in terms of physical variables
\[
 (A^i,\pi_{(A)i} = -\frac{1}{2}{\varepsilon}_{oijk}B^{jk}),
\]
\[
 (B^{oi},\pi_{(B)oi} = B_i),
\]
\begin{equation}
 (C^{ia},\pi_{(C)ia}=-M_{ab}(G^b_{0i}+\frac{1}{2}
 \varepsilon_{oi}{}^{jk}X^b_cG^c_{jk}))
\end{equation}
\[
 (C^{oa},\pi_{(C)oa} = - {\varepsilon}_{ab}B^b),
\]
\[
 (C^{ab},\pi_{(C)ab} = - \frac{1}{2}\varepsilon_{ac}\varepsilon_{bd}
 {\cal D}_0C^{cd}),
\]
 have the algebraic properties
\[
 \{Q^a,\;Q^b\} = 0
\]
 with respect to the Poisson bracket in phase space (5.18). The
 Noether charge operators $\hat{Q}^a$ ($[\hat{Q}^a,\hat{Q}^b]_{+}=0$)
 generate the transformations
\[
 [\hat {A}^i,\hat {Q}^a]=0,\;[\hat{\pi}_{(A)i},\hat{Q}^a]=
 i\{\varepsilon_{0ijk}(\partial^j\hat{C}^{ka}+\hat{A}^j\wedge
 \hat{C}^{ka})+\hat{\pi}_{(C)ib}\wedge\hat {C}^{ab}-
 \varepsilon^{ab}\hat{\pi}_{(C)ob}\wedge\hat {B}_{oi}\},
\]
\[
 [\hat {B}^{0i},\hat {Q}^a]=i\varepsilon^{ab}\hat{\pi}^i_{(C)b},\;
 [\hat {\pi}_{(B)0i},\hat {Q}^a]=-i\varepsilon^{ab}(\partial_i
 \hat\pi_{(C)0b}+\hat {A}_i\wedge\pi_{(C)0b}),
\]
\[
 [\hat {C}^{ib},\hat {Q}^a]_{+}=i(\partial^i \hat {C}^{ab}+
 \hat {A}^i\wedge\hat {C}^{ab}+\varepsilon^{ab}\hat {\pi}_{(B)}^{0i}),
\]
\[
 [\hat {\pi}_{(C)ib},\hat{Q}^a]_{+}=-i\varepsilon_{0ijk}\delta_b^a
 (\partial^j\hat{A}^{k}+\frac{1}{2}\hat{A}^j\wedge\hat{A}^k),
\]
\[
 [\hat {C}^{0b},\hat {Q}^a]_{+}=-i\{\varepsilon^{ab}
 (\partial^i\hat {B}_{0i}+\hat {A}^i\wedge\hat {B}_{0i})+
 2\varepsilon^{ac}\varepsilon^{bd}\hat {\pi}_{(C)cd}\},\;
 [\hat{\pi}_{(C)0b},\hat{Q}^a]_{+}=0.
\]
\[
 [\hat {C}^{bc},\hat {Q}^a]=i\varepsilon^{a\{b}\varepsilon^{c\}d}
 \hat{\pi}_{(C)0d},
\]
\[
 [\hat{\pi}_{(C)bc},\hat{Q}^a]=\frac{i}{2}
 (\partial^i\hat{\pi}_{(C)i\{b}\delta_{c\}}^a+
 \hat {A}^i\wedge\hat{\pi}_{(C)i\{b}\delta_{c\}}^a).
\]
\hspace*{\parindent}
 Next, turn to the analysis of the {\it in}-fields and consider the
 quadratic approximation $S^{(0)}$ for the quantum action (5.17)
\begin{eqnarray}
 S^{(0)}&=&\int\,d^4x\{-\frac{1}{4}{\varepsilon}^{\mu\nu\rho\sigma}
 F_{\mu\nu} B_{\rho\sigma} + \frac{1}{2} A_{\mu}
 A^{\mu}+\nonumber\\&&
 +\frac{1}{2}\varepsilon_{ab}\partial_\mu C_\nu^a
 (\partial^\mu C^{\nu b}-\partial^\nu C^{\mu b})-
 \frac{1}{4}\varepsilon_{ab}\varepsilon_{cd}\partial_\mu
 C^{ac}\partial^\mu C^{bd}+\nonumber\\&&\mbox{}
 +B_\mu\partial_\nu B^{\nu\mu}+\varepsilon_{ab} B^a \partial^\mu C_\mu
 ^b-\frac{1}{2}B_\mu B^\mu-\frac{1}{2}\varepsilon_{ab} B^aB^b\}.
\end{eqnarray}
 The equations of motion for the {\it in}-fields following from Eq.~(5.19)
 are representable in the form
\[
 \Box B_{\mu\nu}=0,\;\;\Box C^a_{\mu}=0,\;\;\Box C^{ab}=0, \nonumber\\
\]
\[
 A_\mu=\frac{1}{2}{\varepsilon}_{\mu\nu\rho\sigma}\partial^{\nu}
 B^{\rho\sigma},\;\;
 B_{\mu}=\partial^{\nu}B_{\nu\mu},\;\;B^{a}=\partial^{\mu}C^a_{\mu}\,.
\]
 The decompositions of $\hat{B}_{\mu\nu}(x)$, $\hat{C}^a_{\mu}(x)$
 in the creation and annihilation operators are given by the relations
 (5.12), (5.13), while in the corresponding decomposition for
 $\hat{C}^{ab}(x)$ one should make allowance for
\begin{eqnarray*}
 (\hat{C}^{ab(-)})^{\dagger}=(-1)^{a+b+1}\hat{C}^{ab(+)}.
\end{eqnarray*}
 The (anti-)commutation relations for the creation and annihilation
 operators read
\[
 {[}\hat{B}^{(-)}_{\mu\nu}(\vec{k}),\hat{B}^{(+)}_{\rho\sigma}(\vec{k'})]
 =\eta_{\mu{[}\rho}\eta_{\sigma{]}\nu}\delta(\vec{k}-\vec{k'}),
\]
\begin{equation}
 {[}\hat{C}^{a(-)}_{\mu}(\vec{k}),\hat{C}^{b(+)}_{\nu}(\vec{k'}){]}_{+}
 ={\varepsilon}^{ab}\eta_{\mu\nu}\delta(\vec{k}-\vec{k'}),
\end{equation}
\[
 {[}\hat{C}^{ab(-)}(\vec{k}),\hat{C}^{cd(+)}(\vec{k'}){]}
 =-{\varepsilon}^{a\{c}{\varepsilon}^{d\}b}\delta(\vec{k}-\vec{k'}).
\]
 Owing to Eq.~(5.20), the doublet of operators $\hat{Q}^a$
\begin{eqnarray*}
 \hat{Q}^{a}&=&\int d^3k\;k^\mu\bigg(
 \hat{B}^{(+)}_{\nu\mu}(\vec{k})\hat{C}^{\nu a(-)}(\vec{k})+
 \hat{C}^{\nu a(+)}(\vec{k})\hat{B}^{(-)}_{\nu\mu})(\vec{k})+\\&&
 \mbox{}
 +\varepsilon_{cb}(\hat{C}^{ab(+)}(\vec{k})\hat{C}_\mu^{c(-)}(\vec{k})+
 \hat{C}_\mu^{c(+)}(\vec{k})\hat{C}^{ab(-)}(\vec{k})\bigg)
\end{eqnarray*}
 generates for the {\it in}-operators the extended BRST symmetry
 transformations
\begin{eqnarray*}
 \delta\hat{B}^{\alpha\beta}&=&-(\partial^\alpha\hat{C}^{\beta a}-
 \partial^\beta\hat{C}^{\alpha a})\mu_a,\\
 \delta\hat{C}^{\alpha a}&=&(\partial^\alpha
 \hat{C}^{ab}-{\varepsilon}^{ab}\hat{B}^{\alpha})\mu_b,\;\;\;
 \delta\hat{B}^\alpha=\partial^\alpha\hat{B}^a\mu_a,\\
 \delta\hat{C}^{ab}&=&\hat{B}^{\{a}{\varepsilon}^{b\}c}\mu_c,
\end{eqnarray*}
 which, for the creation operators, take on the form
\[
 {[}\hat{B}^{(+)}_{\mu\nu}(\vec{k}),\hat{Q}^{a}]
 =k_\mu\hat{C}^{a(+)}_\nu(\vec{k})-k_\nu\hat{C}^{a(+)}_\mu
 (\vec{k}),
\]
\begin{eqnarray}
 {[}\hat{C}^{b(+)}_{\mu}(\vec{k}),\hat{Q}^{a}{]}_{+}
 ={\varepsilon}^{ab}k^\nu\hat{B}^{(+)}_{\nu\mu}(\vec{k})-
 k_\mu \hat{C}^{ab(+)}(\vec{k}),
\end{eqnarray}
\[
 {[}\hat{C}^{bc(+)}(\vec{k}),\hat{Q}^{a}{]}
 =-k^\mu\varepsilon^{a\{b}\hat{C}^{c\}(+)}_\mu(\vec{k}).
\]
 Making use of the decompositions of operators in the local basis
 (5.15) and taking the notation (5.16) into account, we find, by virtue
 of Eq.~(4.21), that the 17 one-particle states of the theory form a
 genuine BRST--antiBRST-singlet
\[
 (\hat{D}^{(+)}_L-\hat{D}^{(+)}_T)|0>,
\]
 two BRST--antiBRST-quartets
 \begin{eqnarray*} (\hat{B}^{(+)}_1+\hat{D}^{(+)}_2)|0>,\;\;
 \hat{Q}^{a}(\hat{B}^{(+)}_1+\hat{D}^{(+)}_2)|0>,\;\;
 \frac{1}{2}\varepsilon_{ab}\hat{Q}^{a}\hat{Q}^{b}(\hat{B}^{(+)}_1+
 \hat{D}^{(+)}_2)|0>,
 \\
 (\hat{B}^{(+)}_2-\hat{D}^{(+)}_1)|0>,\;\;
 \hat{Q}^{a}(\hat{B}^{(+)}_2-\hat{D}^{(+)}_1)|0>,\;\;
 \frac{1}{2}\varepsilon_{ab}\hat{Q}^{a}\hat{Q}^{b}(\hat{B}^{(+)}_2-
 \hat{D}^{(+)}_1)|0>
\end{eqnarray*}
 and a BRST--antiBRST-octet
\begin{eqnarray*}
 \hat{C}^{1(+)}_L|0>,\;\;
 \hat{Q}^{a}\hat{C}^{1(+)}_L|0>,\;\;
 \frac{1}{2}\varepsilon_{ab}\hat{Q}^{a}\hat{Q}^{b}\hat{C}^{1(+)}_L|0>,\\
 \hat{C}^{2(+)}_L|0>,\;\;
 \hat{Q}^{a}\hat{C}^{2(+)}_L|0>,\;\;
 \frac{1}{2}\varepsilon_{ab}\hat{Q}^{a}\hat{Q}^{b}\hat{C}^{2(+)}_L|0>.
\end{eqnarray*}
 Given this, the absense of BRST--antiBRST-singlet pairs and
 BRST--antiBRST-sextets as well as BRST-quartets (antiBRST-singlet
 pairs) and antiBRST-quartets (BRST-singlet pairs) ensures the physical
 unitarity of the theory (see [37]).
%%%%%%%%%%%%%%%%%%%%%%%%%%%%%%%%%%%%%%%%%%%%%%%%%%%%%%%%%%%%%%%%%%%%%%%
\section{Conclusion}
\hspace*{\parindent}
 In this paper, we have studied the unitarity problem for general gauge
 theories within the Lagrangian $Sp(2)$-symmetric scheme proposed by
 Batalin, Lavrov and Tyutin in Refs.~[8--10] and underlied by the
 principle of invariance under the extended BRST transformations. The
 present study is based on the investigation of asymptotic state
 space structure with the help of representations of the algebra of
 generators $\hat{Q}^a$ ($a=1$, 2) of the extended BRST symmetry
 transformations and the ghost charge operator $i\hat{Q}_C$.
 It is shown that the space of representation of the algebra of the
 operators $\hat{Q}^a$, $i\hat{Q}_C$ can be described by 7 types of
 one-particle state complexes referred to in the paper as genuine
 BRST--antiBRST-singlets (physical particles), pairs of
 BRST--antiBRST-singlets, BRST-quartets, antiBRST-quartets,
 BRST--antiBRST-sextets and BRST--antiBRST-octets. The conditions of
 the {\it S}-matrix unitarity are formulated. To provide the unitarity
 of a gauge theory in the Lagrangian $Sp(2)$-symmetric scheme, we
 require that in the theory be no BRST--antiBRST-singlet pairs,
 BRST-quartets, antiBRST-quartets and BRST--antiBRST-sextets. The
 general results are exemplified on the basis of the well-known
 Freedman--Townsend model [34] and the antisymmetric tensor field
 model with auxiliary gauge fields, proposed in Ref.~[32].

\end{document}